\documentclass[11pt]{article}
\usepackage[english]{babel}

\usepackage{amsthm}
\usepackage{amsmath}
\usepackage{graphicx} 
\usepackage[utf8]{inputenc}
\usepackage{subcaption}
\usepackage{tikz}
\usetikzlibrary{positioning}
\usepackage{bm}
\usepackage{multirow}
\usepackage{amssymb}
\usepackage{float}
\usepackage{units}
\usepackage{verbatim}
\usepackage[colorlinks,citecolor=blue,linkcolor=blue,urlcolor=blue,filecolor=blue]{hyperref}

\usepackage{geometry}
\usepackage{thumbpdf,lmodern}

\usepackage{enumitem}
\usepackage{booktabs}

\usepackage{titlesec}

\usepackage[longnamesfirst, round]{natbib}
\setcitestyle{authoryear,open={(},close={)}}

\newcommand{\code}[1]{\texttt{#1}}
\newcommand{\fct}[1]{\code{#1()}}
\newcommand{\proglang}[1]{\code{#1}}
\newcommand{\pkg}[1]{\code{#1}}

\usepackage{fancyvrb}

\makeatletter
\def\th@definition{%
  \thm@notefont{}
  \itshape 
  \thm@headfont{\bfseries}
}
\makeatother

\theoremstyle{definition}
\newtheorem{model}{Model}
\newtheorem{exmp}{Example}[section]

\newenvironment{CodeChunk}{}{}

\DefineVerbatimEnvironment{Code}{Verbatim}{commandchars=\\\{\}}
\DefineVerbatimEnvironment{CodeInput}{Verbatim}{fontshape=it, commandchars=\\\{\}}
\DefineVerbatimEnvironment{CodeOutput}{Verbatim}{commandchars=\\\{\}}

\usepackage{authblk}
  
\newlength{\currentparskip}
\newlength{\currentparindent}
\newenvironment{myabstract}[1]
    {
    \setlength{\currentparskip}{\parskip}
    \setlength{\currentparindent}{\parindent}
    \begin{center}
    \begin{minipage}[]{#1}
    \setlength{\parskip}{\currentparskip}
    \setlength{\parindent}{\currentparindent}
    \small
    \begin{center} \textbf{Abstract} \end{center} \vspace*{-0.5em} \vspace*{-\currentparskip}
    }
    {
    \end{minipage}
    \end{center}
    }
    
\newenvironment{mykeywords}[1]
    {
    \vspace*{-1.5em}
    \begin{center}
    \begin{minipage}[]{#1}
    \small
    \noindent \textbf{Keywords:}
    }
    {
    \end{minipage}
    \end{center}
    }

\begin{document}

\graphicspath{{./img/}} 

\title{\pkg{makemyprior}: Intuitive Construction of Joint Priors for Variance Parameters in \proglang{R}}
\author[1]{Ingeborg Gullikstad Hem\footnote{Corresponding author: \url{ingeborg.hem@ntnu.no}}}
\author[1]{Geir-Arne Fuglstad}
\author[1]{Andrea Riebler}
\affil[1]{Department of Mathematical Sciences, NTNU, Norway}
\date{}
  
\maketitle 

\begin{myabstract}{0.85\linewidth}

\noindent
Priors allow us to robustify inference and to incorporate 
    expert knowledge in Bayesian hierarchical models.
    This is particularly important when there are 
    random effects that are hard to identify based
    on observed data. The challenge
    lies in understanding and controlling the
    joint influence of the priors for the variance
    parameters, and
    \pkg{makemyprior} is an \proglang{R} package that guides the formulation of joint prior distributions
    for variance parameters.
    A joint prior distribution is constructed based on a hierarchical decomposition of the total variance in the model along a tree, and takes the entire model
    structure into account. Users
    input their prior beliefs or express ignorance at each level of the tree. Prior
    beliefs can be general ideas about reasonable ranges of variance values and
    need not be detailed expert knowledge.
    The constructed priors 
    lead to robust inference and guarantee proper posteriors.
    A graphical user interface facilitates construction and assessment of
    different choices of priors through visualization of the tree and joint prior.
    The package aims to expand the toolbox of applied researchers and make priors
    an active component in their Bayesian workflow. 
    
\end{myabstract}

\begin{mykeywords}{0.85\linewidth}
Bayesian hierarchical models,
robust inference,
joint prior distributions, 
hierarchical variance decomposition, 
graphical user interface, 
\proglang{R}.
\end{mykeywords}


\tikzset{
	lab/.style args={[#1]#2}{
    label={[yshift = -48, fill = white, #1]#2}
    },
    mynode/.style={circle, draw = black, fill = black!10, very thick, minimum size = 30},
    mynode2/.style={rounded corners=5, yshift = 15, fill = black!10, minimum height = 15},
    mynode3/.style={rounded corners=5, yshift = 15, fill = black!10, minimum width = 25, minimum height = 20}
}

\tikzset{
theory_tree1/.pic = {
	\node[mynode2] (topnode) [] {$a+b+c+\varepsilon$};
	\node[mynode2] (residuals) [below = of topnode, xshift = 20] {$\varepsilon$};
	\node[mynode2] (node123) [below = of topnode, xshift = -20] {$a+b+c$};
	\node[mynode2] (node1) [below = of node123, xshift = -20] {$a$};
	\node[mynode2] (node2) [below = of node123] {$b$};
	\node[mynode2] (node3) [below = of node123, xshift = 20] {$c$};

	\draw[->] (topnode) -- (node123);
	\draw[->] (topnode) -- (residuals);
	\draw[->] (node123) -- (node1);
	\draw[->] (node123) -- (node2);
	\draw[->] (node123) -- (node3);
},
theory_tree2/.pic = {
	\node[mynode2] (topnode) [] {$a+b+c+\varepsilon$};
	\node[mynode2] (residuals) [below = of topnode, xshift = 20] {$\varepsilon$};
	\node[mynode2] (node123) [below = of topnode, xshift = -20] {$a+b+c$};
	\node[mynode2] (node3) [below = of node123, xshift = 20] {$c$};
	\node[mynode2] (node12) [below = of node123, xshift = -20] {$a+b$};
	\node[mynode2] (node1) [below = of node12, xshift = -20] {$a$};
	\node[mynode2] (node2) [below = of node12, xshift = 20] {$b$};

	\draw[->] (topnode) -- (node123);
	\draw[->] (topnode) -- (residuals);
	\draw[->] (node123) -- (node12);
	\draw[->] (node123) -- (node3);
	\draw[->] (node12) -- (node1);
	\draw[->] (node12) -- (node2);
},
theory_tree3/.pic = {
	\node[mynode2] (residuals) [xshift = -20] {$\varepsilon$};
	\node[mynode2] (node123) [xshift = 20] {$a+b+c$};
	\node[mynode2] (node1) [below = of node123, xshift = -20] {$a$};
	\node[mynode2] (node2) [below = of node123] {$b$};
	\node[mynode2] (node3) [below = of node123, xshift = 20] {$c$};

	\draw[->] (node123) -- (node1);
	\draw[->] (node123) -- (node2);
	\draw[->] (node123) -- (node3);
},
theory_tree4/.pic = {
	\node[mynode2] (node1) [] {$a$};
	\node[mynode2] (node2) [right = of node1, xshift = -20, yshift = -15] {$b$};
	\node[mynode2] (node3) [right = of node2, xshift = -20, yshift = -15] {$c$};
	\node[mynode2] (residuals) [right = of node3, xshift = -20, yshift = -15] {$\varepsilon$};
},
theory_tree5/.pic = {
	\node[mynode2] (topnode) [] {$a+b+c+\varepsilon$};
	\node[mynode2] (node1) [below = of topnode, xshift = -30] {$a$};
	\node[mynode2] (node2) [below = of topnode, xshift = -10] {$b$};
	\node[mynode2] (node3) [below = of topnode, xshift = 10] {$c$};
	\node[mynode2] (residuals) [below = of topnode, xshift = 30] {$\varepsilon$};

	\draw[->] (topnode) -- (residuals);
	\draw[->] (topnode) -- (node1);
	\draw[->] (topnode) -- (node2);
	\draw[->] (topnode) -- (node3);
}
}

\tikzset{
software_tree1/.pic = {
	\node[mynode2] (topnode) [] {$a+b+\varepsilon$};
	\node[mynode2] (residuals) [below = of topnode, xshift = 20] {$\varepsilon$};
	\node[mynode2] (node1) [below = of topnode, xshift = -20] {$a$};
	\node[mynode2] (node2) [below = of topnode, xshift = 0] {$b$};

	\draw[->] (topnode) -- (residuals);
	\draw[->] (topnode) -- (node1);
	\draw[->] (topnode) -- (node2);
},
software_tree2/.pic = {
	\node[mynode2] (topnode) [] {$a+b+\varepsilon$};
	\node[mynode2] (residuals) [below = of topnode, xshift = 20] {$\varepsilon$};
	\node[mynode2] (node12) [below = of topnode, xshift = -20] {$a+b$};
	\node[mynode2] (node1) [below = of node12, xshift = -20] {$a$};
	\node[mynode2] (node2) [below = of node12, xshift = 20] {$b$};

	\draw[->] (topnode) -- (node12);
	\draw[->] (topnode) -- (residuals);
	\draw[->] (node12) -- (node1);
	\draw[->] (node12) -- (node2);
},
software_tree3/.pic = {
	\node[mynode2] (node1) [] {$a$};
	\node[mynode2] (node2) [right = of node1, xshift = -20, yshift = -15] {$b$};
	\node[mynode2] (residuals) [right = of node2, xshift = -20, yshift = -15] {$\varepsilon$};
},
software_tree4/.pic = {
	\node[mynode2] (residuals) [xshift = 20] {$\varepsilon$};
	\node[mynode2] (node12) [xshift = -20] {$a+b$};
	\node[mynode2] (node1) [below = of node12, xshift = -20] {$a$};
	\node[mynode2] (node2) [below = of node12, xshift = 20] {$b$};

	\draw[->] (node12) -- (node1);
	\draw[->] (node12) -- (node2);
}
}

\tikzset{
latin_tree1/.pic = {
	\node[mynode3] (node1) [] {$a$};
	\node[mynode3] (node2) [right = of node1, xshift = -20, yshift = -15] {$b$};
	\node[mynode3] (node3) [right = of node2, xshift = -20, yshift = -15] {$c^{(1)}$};
	\node[mynode3] (node4) [right = of node3, xshift = -20, yshift = -15] {$c^{(2)}$};
	\node[mynode3] (residuals) [right = of node4, xshift = -20, yshift = -15] {$\varepsilon$};
},
latin_tree2/.pic = {
	\node[mynode3] (topnode) [] {$a + b + c^{(1)} + c^{(2)} + \varepsilon$};;
	\node[mynode3] (node1) [below = of topnode, xshift = -68] {$a$};
	\node[mynode3] (node2) [below = of topnode, xshift = -34] {$b$};
	\node[mynode3] (node3) [below = of topnode, xshift = 0] {$c^{(1)}$};
	\node[mynode3] (node4) [below = of topnode, xshift = 34] {$c^{(2)}$};
	\node[mynode3] (residuals) [below = of topnode, xshift = 68] {$\varepsilon$};

	\draw[->] (topnode) -- (residuals);
	\draw[->] (topnode) -- (node1);
	\draw[->] (topnode) -- (node2);
	\draw[->] (topnode) -- (node3);
	\draw[->] (topnode) -- (node4);
},
latin_tree3/.pic = {
	\node[mynode3] (topnode) [] {$a + b + c^{(1)} + c^{(2)} + \varepsilon$};
	\node[mynode3] (residuals) [below = of topnode, xshift = 35] {$\varepsilon$};
	\node[mynode3] (node123) [below = of topnode, xshift = -35] {$a + b + c^{(1)} + c^{(2)}$};
	\node[mynode3] (node1) [below = of node123, xshift = -35] {$a$};
	\node[mynode3] (node2) [below = of node123, xshift = -5] {$b$};
	\node[mynode3] (node3) [below = of node123, xshift = 41] {$c^{(1)} + c^{(2)}$};
	\node[mynode3] (node31) [below = of node3, xshift = -18] {$c^{(1)}$};
	\node[mynode3] (node32) [below = of node3, xshift = 18] {$c^{(2)}$};

	\draw[->] (topnode) -- (node123);
	\draw[->] (topnode) -- (residuals);
	\draw[->] (node123) -- (node1);
	\draw[->] (node123) -- (node2);
	\draw[->] (node123) -- (node3);
	\draw[->] (node3) -- (node31);
	\draw[->] (node3) -- (node32);
}
}

\tikzset{
genetic_tree1/.pic = {
	\node[mynode2] (node1) [] {$a$};
	\node[mynode2] (node2) [right = of node1, xshift = -20, yshift = -15] {$d$};
	\node[mynode2] (node3) [right = of node2, xshift = -20, yshift = -15] {$x$};
	\node[mynode2] (residuals) [right = of node3, xshift = -20, yshift = -15] {$\varepsilon$};
},
genetic_tree2/.pic = {
	\node[mynode2] (topnode) [] {$a+d+x+\varepsilon$};
	\node[mynode2] (node1) [below = of topnode, xshift = -30] {$a$};
	\node[mynode2] (node2) [below = of topnode, xshift = -10] {$d$};
	\node[mynode2] (node3) [below = of topnode, xshift = 10] {$x$};
	\node[mynode2] (residuals) [below = of topnode, xshift = 30] {$\varepsilon$};

	\draw[->] (topnode) -- (residuals);
	\draw[->] (topnode) -- (node1);
	\draw[->] (topnode) -- (node2);
	\draw[->] (topnode) -- (node3);
},
genetic_tree3/.pic = {
	\node[mynode2] (topnode) [] {$a+d+x+\varepsilon$};
	\node[mynode2] (residuals) [below = of topnode, xshift = 20] {$\varepsilon$};
	\node[mynode2] (node123) [below = of topnode, xshift = -20] {$a+d+x$};
	\node[mynode2] (node3) [below = of node123, xshift = -20] {$a$};
	\node[mynode2] (node12) [below = of node123, xshift = 20] {$d+x$};
	\node[mynode2] (node1) [below = of node12, xshift = -20] {$d$};
	\node[mynode2] (node2) [below = of node12, xshift = 20] {$x$};

	\draw[->] (topnode) -- (node123);
	\draw[->] (topnode) -- (residuals);
	\draw[->] (node123) -- (node12);
	\draw[->] (node123) -- (node3);
	\draw[->] (node12) -- (node1);
	\draw[->] (node12) -- (node2);
}
}

\tikzset{
kenya_tree1/.pic = {
	\node[mynode2] (topnode) [] {$u+v+\nu$};
	\node[mynode2] (uv) [below = of topnode, xshift = -20] {$u+v$};
	\node[mynode2] (u) [below = of uv, xshift = -20] {$u$};
	\node[mynode2] (v) [below = of uv, xshift = 20] {$v$};
	\node[mynode2] (nu) [below = of topnode, xshift = 20] {$\nu$};

	\draw[->] (topnode) -- (nu);
	\draw[->] (topnode) -- (uv);
	\draw[->] (uv) -- (u);
	\draw[->] (uv) -- (v);
},
kenya_tree2/.pic = {
	\node[mynode2] (topnode) [] {$u+v+\nu+\varepsilon$};
	\node[mynode2] (eps) [below = of topnode, xshift = 20] {$\varepsilon$};
	\node[mynode2] (uvnu) [below = of topnode, xshift = -20] {$u+v+\nu$};
	\node[mynode2] (uv) [below = of uvnu, xshift = -20] {$u+v$};
	\node[mynode2] (u) [below = of uv, xshift = -20] {$u$};
	\node[mynode2] (v) [below = of uv, xshift = 20] {$v$};
	\node[mynode2] (nu) [below = of uvnu, xshift = 20] {$\nu$};

	\draw[->] (topnode) -- (eps);
	\draw[->] (topnode) -- (uvnu);
	\draw[->] (uvnu) -- (uv);
	\draw[->] (uvnu) -- (nu);
	\draw[->] (uv) -- (u);
	\draw[->] (uv) -- (v);
}
}

\section{Introduction}
\label{sec:intro}





Bayesian modelling is more available than
ever
through fast and easy-to-use software programs for Bayesian inference such as
Integrated Nested Laplace Approximations 
\citep[INLA,][]{rue2009} in \proglang{R} 
through \pkg{INLA} (see \url{www.r-inla.org}),
\proglang{Stan} \citep{carpenter2017} with the \proglang{R} interface \pkg{rstan} \citep{rstan2020}
and dependencies such as
\pkg{rstanarm} \citep{rstanarm2020}, \pkg{brms} \citep{brms},
\pkg{shinystan} \citep{shinystan2020} and \pkg{loo} \citep{loo2020}, WinBUGS \citep{lunn-etal-2000} with the \proglang{R} interface R2WinBUGS \citep{r2winbugs}, OpenBUGS \citep{openbugs} with the \proglang{R} interface R2OpenBUGS \citep{r2openbugs}, 
Template Model Builder \citep[\pkg{TMB},][]{tmb}, 
\pkg{JAGS} \citep{jags_manual}, 
Bayesian Analysis Toolkit \citep[\pkg{BAT},][]{caldwell2009}, 
and more.
These programs offer many ways to
construct complex models suitable for a wide range of applications.
Default settings for priors and hyperparameters are usually given, and instructions
on how to change the default settings can be found in the software documentation.
However, 
even though there is an increasing
focus on that prior distributions should be chosen consciously
\citep{zondervan2017, gelman2020workflow, smid2020},
guides on how the priors should be chosen are missing.
Our goal with \pkg{makemyprior} 
is to close this gap and to increase the awareness of prior choices.
We empower users to actively select 
priors that are suitable for their model structure and application at hand.
The user is made aware of which priors are used in the model and what they express.
In particular, the assumptions underlying the default settings are made clear.
The package is available from
the Comprehensive \proglang{R} Archive Network (CRAN):
\url{https://CRAN.R-project.org/package=makemyprior}.


In Bayesian hierarchical models, the variation in
the observations is modelled through a combination of an observation model,
a latent model, and priors for the parameters. The observation model defines a
generative model for the observed data conditional
on the latent model. The link between the latent model and each
observation acts through the transformation of a linear predictor, 
which we assume is a linear combination of fixed and random effects.
The goal of the linear predictor is to explain the variation in
the true signal.
We concentrate on
latent Gaussian models, where the linear predictor is 
composed of random model components that 
follow multivariate Gaussian distributions conditional on the parameters.
These parameters control the random effects and need prior distributions.
We 
focus on
the most central type of model parameters: the variance parameters.


Parameters controlling means and medians, such as the coefficients of fixed effects,
are close to the data and tolerate vague priors \citep{goel1981, gelman2020workflow}.
We 
give them
Gaussian priors with zero mean and a fixed high variance.
The specification of a prior distribution for a variance parameter
is challenging 
\citep{lambert2005, gelman2017prior},
but is at the same time a strong feature of Bayesian inference.
Prior knowledge, obtained from previous experiments or comparable investigations, and expert knowledge can be included to make the model more robust.
It is convenient to use the default priors included in programs for inference, 
however,
by using defaults we are not fully utilizing
the Bayesian framework.
Currently,
there are no other \proglang{R} packages that intuitively allows the proper inclusion of such knowledge and 
simple visualization of chosen priors in a straightforward manner. 

The \pkg{makemyprior} package applies the hierarchical decomposition (HD) prior framework proposed by \citet{fuglstad2020},
where the variation
in the observed data 
is distributed 
to the random effects following a \textit{prior tree structure}.
In this tree, the leaf nodes represent the random effects specified in the linear predictor,
and the top (root) node represents the sum of the random effect variance, denoted the 
\textit{total variance}.
How much variance is distributed from a parent 
node to its child nodes is determined by a 
split in the tree, and 
this procedure continues down until the leaf nodes.
This gives us a parameterization with \textit{proportions of variance}, instead of the more
common variance parameter parameterization.
The priors for those variance proportion parameters can be specified intuitively and transparently as they often coincide with the scale on which prior or expert knowledge exists,
such as in genomic modelling \citep{holand2013, hem2021} and disease mapping \citep{wakefield2006}.
Through the \pkg{makemyprior} package, we
raise the awareness for prior selection and 
guide the user to formulate sensible prior distributions, that are automatically computed and
visualized in an intuitive manner. 

Prior distribution can be either specified directly within \proglang{R}
or through a graphical user interface (GUI). 
In the GUI the user can inspect the prior tree and adapt it as needed. 
One can click through the splits independently to specify the beliefs for each split. 
A guide asking relevant questions about the model to ease the prior
specification process further is also available.
The user can be ignorant and distribute the variance equally to the child nodes through a Dirichlet prior, or exploit expert knowledge implemented via penalized complexity 
\citep[PC,][]{simpson2017} priors.
After completing the prior specification, inference
can be carried out directly
with the \proglang{R} packages
\pkg{rstan} and \pkg{INLA}.

We begin with 
explaining the concepts of total variance
and hierarchical variance decomposition 
through two motivating examples in Section \ref{sec:motivating}.
We then introduce the necessary background in
Section \ref{sec:theory}
before we present
the \pkg{makemyprior} package
with general explanations on how to use it in Section \ref{sec:software}.
Section \ref{sec:examples} gives more detailed examples showing how to use the package in various
situations.
A summary and discussion is given in Section \ref{sec:summary}.


\section{Motivating examples}
\label{sec:motivating}


We demonstrate the core
ideas of total variance and hierarchical decomposition of the
total variance through two illustrative examples. 
These examples are simplified versions of examples we use 
in Section \ref{sec:examples}.

One of the key quantities in quantitative genetics
concerns the
distribution of observed variation to genetic and environmental 
sources.
In this setting, good intuition exists on the ratio of genetic to phenotypic variation, also known as 
the 
\textit{heritability}, whereas it is more difficult to define suitable priors separately on the two variance parameters.

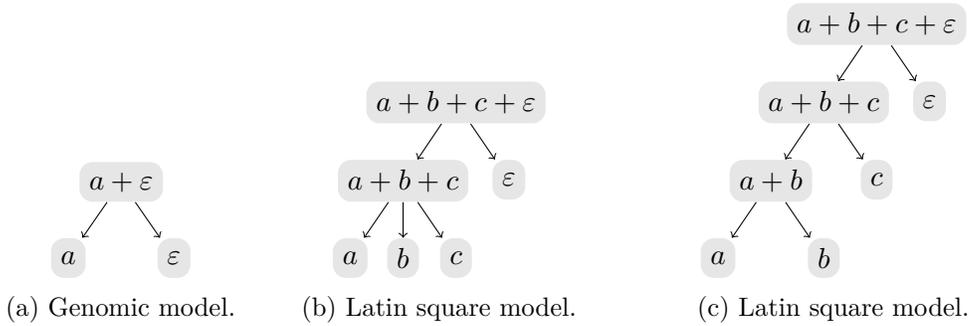
\begin{figure}[t!]
\centering
\begin{subfigure}[t]{0.2\textwidth}
    \centering
    \begin{tikzpicture}[]
    	\node[mynode2] (topnode) [] {$a+\varepsilon$};
    	\node[mynode2] (residuals) [below = of topnode, xshift = 20] {$\varepsilon$};
    	\node[mynode2] (node123) [below = of topnode, xshift = -20] {$a$};
    
    	\draw[->] (topnode) -- (node123);
    	\draw[->] (topnode) -- (residuals);
    	
    \end{tikzpicture}
    \caption{Genomic model.}
    \label{fig:theory:trees_a}
\end{subfigure}
\begin{subfigure}[t]{0.33\textwidth}
    \centering
    \begin{tikzpicture}[]
    	\node[mynode2] (topnode) [] {$a+b+c+\varepsilon$};
    	\node[mynode2] (residuals) [below = of topnode, xshift = 20] {$\varepsilon$};
    	\node[mynode2] (node123) [below = of topnode, xshift = -20] {$a+b+c$};
    	\node[mynode2] (node1) [below = of node123, xshift = -20] {$a$};
    	\node[mynode2] (node2) [below = of node123] {$b$};
    	\node[mynode2] (node3) [below = of node123, xshift = 20] {$c$};
    
    	\draw[->] (topnode) -- (node123);
    	\draw[->] (topnode) -- (residuals);
    	\draw[->] (node123) -- (node1);
    	\draw[->] (node123) -- (node2);
    	\draw[->] (node123) -- (node3);
    	
    \end{tikzpicture}
    \caption{Latin square model.}
    \label{fig:theory:trees_b}
\end{subfigure}
\begin{subfigure}[t]{0.33\textwidth}
    \centering
    \begin{tikzpicture}[]
    	\node[mynode2] (topnode) [] {$a+b+c+\varepsilon$};
    	\node[mynode2] (residuals) [below = of topnode, xshift = 20] {$\varepsilon$};
    	\node[mynode2] (node123) [below = of topnode, xshift = -20] {$a+b+c$};
    	\node[mynode2] (node12) [below = of node123, xshift = -20] {$a+b$};
    	\node[mynode2] (node1) [below = of node12, xshift = -20] {$a$};
    	\node[mynode2] (node2) [below = of node12, xshift = 20] {$b$};
    	\node[mynode2] (node3) [below = of node123, xshift = 20] {$c$};
    
    	\draw[->] (topnode) -- (node123);
    	\draw[->] (topnode) -- (residuals);
    	\draw[->] (node123) -- (node12);
    	\draw[->] (node12) -- (node1);
    	\draw[->] (node12) -- (node2);
    	\draw[->] (node123) -- (node3);
    	
    \end{tikzpicture}
    \caption{Latin square model.}
    \label{fig:theory:trees_c}
\end{subfigure}
\caption{%
    Prior trees for (\subref{fig:theory:trees_a}) the genomic model in Example \ref{exmp:genomic}, and (\subref{fig:theory:trees_b}, \subref{fig:theory:trees_c}) the latin square model in Example \ref{exmp:doe}.
    }
\label{fig:theory:trees}
\end{figure}

\begin{exmp}[Genomic models]
\label{exmp:genomic}

Consider a group of $n$ individuals, where each individual $i$ has
an observed phenotype $y_i$. A simple genomic model is:
\begin{equation}
    y_i = \mu + a_i + \varepsilon_i, \quad i = 1, \dots, n, 
    \label{eq:theory:genomic}
\end{equation}
where $\mu$ is an intercept,
$\bm{a} = (a_1, \dots, a_{n})^\top \sim \mathcal{N}_{n}(\bm{0}, \sigma_{\mathrm{a}}^2 \mathbf{A})$ is
an additive genetic effect, and 
$\bm{\varepsilon} = (\varepsilon_1, \dots, \varepsilon_{n})^\top \sim \mathcal{N}_{n}(0, \sigma_{\mathrm{\varepsilon}}^2\mathbf{I}_{n})$
is environmental noise. 
The covariance matrix $\mathbf{A}$ is calculated based
on genetic sequencing of the $n$ individuals and is scaled so that $\sigma_{\mathrm{a}}^2$ is representative of the variance arising
from the genetic effect; see \citet{selle2019, hem2021} for
details.

In this simple model, two key quantities are 
the 
total variance
$\sigma_{\mathrm{P}}^2 = \sigma_\mathrm{a}^2 +\sigma_{\varepsilon}^2$, 
also known as the \emph{phenotypic variance},
and the heritability
$h^2 = \sigma_{\mathrm{a}}^2/(\sigma_\mathrm{a}^2+\sigma_\varepsilon^2)$,
which is the proportion of the phenotypic variance explained 
by the genetic effect. When expert knowledge is available
about these two quantities, this information
can be directly exploited through a joint prior assigned to
phenotypic variance and heritability. 
A simple and intuitive visualization of 
this parameterization is given by the tree in
Figure \ref{fig:theory:trees_a} where the phenotypic variance
$\sigma_{\mathrm{P}}^2 = \sigma_\mathrm{a}^2 +\sigma_{\varepsilon}^2$ 
in the top (root) node
is distributed to the additive genetic variance
$\sigma_\mathrm{a}^2$ and the environmental variance $\sigma_{\varepsilon}^2$ in the two leaf nodes. 
Note that 
the intercept is treated independently with a vague Gaussian prior.
See 
\citet{hem2021} for a detailed description.

\end{exmp}

The idea of expressing a parameterization that is given in terms of
total variance and proportions of variances through a tree, extends to more complex models with more random effects.
For example, when analysing data arising from designed experiments.

\begin{exmp}[Latin square design]
\label{exmp:doe}

Based on the ideas in \cite{fuglstad2020}, we assume that an agricultural 
field is split into rows and columns resulting in a $9 \times 9$
grid, and that one out of $9$ different strengths of fertilizer
is applied in each grid cell. Outcomes $y_{i,j}$ are observed in
row $i$ and column $j$ under treatment $k[i,j]$, and modelled through
\begin{equation}
    y_{i,j} = \alpha + \beta \cdot k[i,j] + a_i + b_j + c_{k[i,j]} + \varepsilon_{i,j}, \quad i,j = 1, \dots, 9,
    \label{eq:theory:latin}
\end{equation}
where $\alpha$ is an intercept and $\beta \cdot k[i,j]$ is a linear effect of treatment. 
$\alpha$ and $\beta$ are assigned vague Gaussian priors $\mathcal{N}(0, \sigma = 1000)$.
$\bm{a} = (a_1, \dots, a_9)^\top \sim 
\mathcal{N}_9(\bm{0}, \sigma_{\mathrm{a}}^2 \mathbf{I}_9)$
is a row effect,
$\bm{b} = (b_1, \dots, b_9)^\top \sim 
\mathcal{N}_9(\bm{0}, \sigma_{\mathrm{b}}^2 \mathbf{I}_9)$
is a column effect, 
$\bm{c} = (c_1, \dots, c_9)^\top \sim 
\mathcal{N}_9(\bm{0}, \sigma_{\mathrm{c}}^2 \mathbf{I}_9)$ is a treatment effect, and the
residual noise is
$\bm{\varepsilon} = (\varepsilon_{1,1},\varepsilon_{1,2} \dots, \varepsilon_{9,9})^\top \sim \mathcal{N}_{81}(\bm{0}, \sigma_{\varepsilon}^2 \mathbf{I}_{81})$. 
$\bm{a}$, $\bm{b}$ and $\bm{c}$ have sum-to-zero constrains.
The individual variances, $\sigma_{\mathrm{a}}^2$, 
$\sigma_{\mathrm{b}}^2$, $\sigma_{\mathrm{c}}^2$
and $\sigma_{\varepsilon}^2$, are nuisance
parameters, and it may be difficult to have prior knowledge about them.

Figures
\ref{fig:theory:trees_b} and \ref{fig:theory:trees_c} visualize
two ways in which the total variance
$\sigma_{\mathrm{a+b+c}+\varepsilon}^2 = \sigma_{\mathrm{a}}^2 + 
\sigma_{\mathrm{b}}^2 + \sigma_{\mathrm{c}}^2 +
\sigma_{\varepsilon}^2$ can be distributed to
the individual variances $\sigma_{\mathrm{a}}^2$, 
$\sigma_{\mathrm{b}}^2$, $\sigma_{\mathrm{c}}^2$
and $\sigma_\varepsilon^2$ in the leaf nodes. 
Using Figure \ref{fig:theory:trees_b},
we could envision that, in the top split, shrinkage is applied to
the latent variance 
$\sigma_{\mathrm{a+b+c}}^2 = \sigma_{\mathrm{a}}^2+\sigma_{\mathrm{b}}^2+\sigma_{\mathrm{c}}^2$ 
relative to the residual variance $\sigma_\varepsilon^2$ with the goal of
reducing the risk of overfitting. 
Then in the second split, we
could envision that we want to express ignorance about how the
latent variance is distributed to
the individual variances $\sigma_{\mathrm{a}}^2$, $\sigma_{\mathrm{b}}^2$ and
$\sigma_{\mathrm{c}}^2$. Alternatively, using Figure \ref{fig:theory:trees_c},
we may want to express ignorance about how 
$\sigma_{\mathrm{a+b}}^2 = \sigma_{\mathrm{a}}^2 + \sigma_{\mathrm{b}}^2$ 
is distributed to 
$\sigma_{\mathrm{a}}^2$ and $\sigma_{\mathrm{b}}^2$, but apply shrinkage to the 
variance of the treatment variance $\sigma_{\mathrm{c}}^2$ relative to
$\sigma_{\mathrm{a+b}}^2$. The full details can be found in
\citet{fuglstad2020}.

\end{exmp}

Examples \ref{exmp:genomic} and \ref{exmp:doe} make it clear that
in some cases it is natural to think in terms of proportions of
variances. However, explicitly writing out how the proportions
are defined may obfuscate the key ideas that one want to express. Therefore, trees such as shown in Figure \ref{fig:theory:trees}
are critical to define such priors. 

We want to emphasize that a prior can be chosen in many ways,
and there are no wrong choices.
Our message is that 
it is a choice to use the default prior, it is a choice
to use literature-based priors, 
and it is a choice to use a prior based on prior and expert knowledge.
We believe it is important to communicate this
and to make prior selection an active part of the Bayesian workflow.

\section{Background}
\label{sec:theory}

In this section, we present the necessary theoretical and 
methodological 
background
behind the \pkg{makemyprior} package
and introduce terminology used throughout
the paper.

\subsection{Hierarchical variance decomposition along a prior tree}
\label{sec:theory:simple}

This section introduces all information needed to use the \pkg{makemyprior} 
package through the guide provided in the graphical user interface.
We include more specific details about the different priors in Section 
\ref{sec:theory:details}.

\subsubsection{General definition of a prior tree}

A prior tree is a directed acyclic graph consisting of a \textit{top node}, \textit{split nodes} and \textit{leaf nodes}, connected by directed edges.
Each random model component is represented by a leaf node.
The top node in the tree represents total variance, i.e., the sum of the variance of the 
leaf nodes (random effects).
We follow
\citet{fuglstad2020} and consider variance that is not explained by fixed effects,
and therefore omit fixed effects from the 
hierarchical decomposition (HD) prior and the prior tree. 
Two or more leaf nodes are combined
in a split node in a way that reflects the hierarchical
model and our prior beliefs 
about how the total variance is distributed.
The split nodes represent a variance proportion.

In some cases we only want to include a subset of the model
components in the same prior tree, and in that way have several trees. 
An example is the animal model \citep[e.g.,][]{holand2013}, a
mixed model that is usually used to decompose environmental and genetic
variances in animal populations. 
The genetic contribution might be split into several sub-components, like
additive, dominance and mutational variance. This is
an extension of the genomic model in Example \ref{exmp:genomic}.
We may have a good intuition on the
absolute magnitude of the variation that is explained by the environment
and by all the genetic contributions separately, but only have
knowledge about the relative magnitude of the genetic variation. 
The genetic contributions are often confounded, and it is useful to
split the variance of the genetic contribution using a prior tree.
In that way we have both variance parameters, total variance parameters, and variance proportions.

If a variance component is assigned an individual prior, its corresponding random effect is represented by a
\textit{singleton}: a node not connected to any other nodes. The singletons can be considered as trees with only one node.
Several prior trees gives a prior forest. We refer
to the forest of trees as the \textit{prior tree structure} of the model.
Each tree in a tree structure is associated with their own joint prior distribution, independent of the priors belonging
to the other trees.
The tree structure is made based on prior knowledge about the
model, data and problem at hand.

\subsubsection{Defining priors for the split nodes: Shrinkage versus ignorance}

Given a tree structure describing the distribution of the variance in the
model, we can use the rest of our pre-existing knowledge to steer the variance to the different model components 
by choosing suitable priors for the different parameters belonging to the prior tree.

A good prior distribution can improve the robustness of the inference, by helping to
avoid estimating spurious effects.
We apply the penalized complexity (PC) priors 
of \citet{simpson2017}
as they shrink towards a so-called base model and are thus robust by design.
For a random effect with zero mean and variance parameter 
$\sigma^2$, the base model is
a model where $\sigma^2 = 0$, meaning the effect does not contribute in the model.
This gives a simpler model with one effect less.
When the parameter varies, we move away from the base model, and this
deviation is penalized, ensuring that we do not overfit the model.
The hyperparameters are chosen using prior knowledge on an interpretable
scale relevant for the parameter. For example, a tail probability for a
standard deviation.

Any model parameter can be assigned a PC prior.
For a variance proportion parameter, the base model can be any value
in $[0, 1]$. In the corner cases $0$ and $1$, we are in a similar situation
as for the variance parameter, 
where the base model represents a model
with fewer random effects as one of them shrinks away.
The heritability $h^2$ in Example \ref{exmp:genomic} is a variance proportion.
We can imagine that a geneticist 
says
that the heritability is around $0.4$ for this phenotype and species, 
but that it is unsure whether
the additive effect contributes or not. Then we can use a PC prior
that shrinks the heritability to $0$, which
gives only residual effect in the base model,
and ensure that the prior has a median of $0.4$, 
i.e., 
$\text{Prob}(h^2 > 0.4) = 0.5$.

If the base model is somewhere between $0$ and
$1$, we express that both 
effects
involved in the variance proportion should be present in the model.
If the geneticist is certain that the additive effect is present in the model,
we can choose a PC prior that shrinks the heritability $h^2$ towards $0.4$, and
with median at the same value: $\text{Prob}(h^2 > 0.4) = 0.5$.
This corresponds to
$40$\% additive and $60$\% residual effect in the base model.
See Section \ref{sec:theory:details} and \citet{fuglstad2020} for details.

By using a \textit{multi-split}, a split with more than two child nodes, the user
expresses no strong opinions on how the variance is distributed among the components
involved in the split.
Then we use a prior that assigns an equal amount of variance to each of the 
model effects involved in the split through
the ignorant symmetric Dirichlet prior. 
This ignorant prior can also be used on a dual split, in which case it
reduces to a uniform prior on $[0, 1]$, and 
with that the user expresses that
no prior information is used in that split.

\subsubsection{Defining priors for top nodes and singletons}

Appropriate prior distributions for variance parameters varies with the likelihood.
We describe the train of thought when specifying prior distributions for 
variance parameters when the likelihood is Gaussian, binomial and Poisson.

In a model with a Gaussian likelihood, the total variance is usually easy to 
identify and does not need an informative
prior.
This is a parameter close to the data in the model and
settles with a vague prior \citep{goel1981, gelman2020workflow}.
For the total variance in the top nodes,
\cite{fuglstad2020} recommend the scale-invariant, improper Jeffreys' prior
when all model components are involved in one single tree.
This prior does not require any hyperparameters and is straight-forward to use.

In cases where the user has specific knowledge
about the total variance, a proper prior can be used to include
this knowledge in the prior.
In cases where the nodes are not all involved in the same tree 
and we have a prior forest,
a scale-invariant prior is not meaningful, and an improper prior may lead
to an improper posterior.
This implies that
singletons always must be assigned a proper prior.
Due to its desirable shrinking properties
\cite{fuglstad2020} recommend the PC prior \citep{simpson2017} 
for the variances, where 
the hyperparameters can be selected 
using a tail probability of the standard deviation:
$\mathrm{Prob}(\sigma > U) = \alpha$. We denote this prior 
$\mathrm{PC}_{\mathrm{0}}(U, \alpha)$, where the subscript indicates
that the shrinkage is towards $0$.
For the genomic model in Example \ref{exmp:genomic}, assume we have
prior or expert 
knowledge saying it is unlikely that the total variance 
in the observed data is greater than $4$. 
We want to use this knowledge, and choose
$U = \sqrt{4}$. 
The value we choose for $\alpha$ says something about how certain we
are in the value of $U$. Using $\alpha = 0.05$ is a suitable choice,
so that $\mathrm{Prob}(\sigma > \sqrt{4}) = 0.05$
and $\sigma \sim \mathrm{PC}_{\mathrm{0}}(\sqrt{4}, 0.05)$.

Other likelihoods require proper priors on all variances, also the total variance,
as scale-invariance is not meaningful for data that are not Gaussian.
Again we
follow the recommendation of \citet{fuglstad2020} and suggest PC priors.
However, instead of choosing a prior using the upper tail probability of the standard deviation, 
we think on a different scale than for Gaussian data, and
choose a credible interval for the variance parameter on a suitable scale. 
Then we
transform the interval into an upper tail probability
$\mathrm{Prob}(\sigma > U) = \alpha$.
For both binomial likelihood with logit link and Poisson likelihood with
log link, 
an exponential scale is appropriate.
This will correspond to thinking on 
odds-ratio scale for binomial data, and on
the scale of the data for Poisson data.

Assume we have a model with linear predictor $\eta_i = \mu + a_i + b_i$
and binomial likelihood with logit link function.
An intuitive
way of choosing a prior for the total variance of $a_i + b_i$ 
is to choose an equal-tailed 
credible interval 
for the effect of the random effects on the odds-ratio, 
$\exp(a_i + b_i)$, i.e.,
$\mathrm{Prob}(l < \exp(a_i + b_i) < u) = p$ \citep{fong2010}.
For example, we can say we want 
$\mathrm{Prob}(0.1 < \exp(a_i + b_i) < 10) = 0.9$.
This corresponds to a 90\% credible interval $[0.1, 10]$ for
$\exp(a_i + b_i)$.
The idea is the same for a Poisson likelihood with a log link: We can think on the effect
of the random effects on the relative risk.


\subsection{Priors for variance proportions}
\label{sec:theory:details}

Details around the prior distributions discussed in the previous section
are presented here. While the \pkg{makemyprior}
package can be used without knowing these details, they are needed 
to fully understand the hierarchical decomposition (HD) prior framework.

\subsubsection{Penalized complexity}

The penalized complexity (PC) priors of \citet{simpson2017} are
based on principles independent of model and application.
They use the
\textit{distance} $d(\cdot)$ between the base model and a flexible extension
of the base model, measured using 
the Kullback-Leibler divergence.
In the base model the parameter of interest $\theta$ is fixed to $\theta_0$. 
In
the flexible model 
$\theta$ is allowed to vary,
and the PC prior induces shrinkage towards the base model, which gives a robust prior that
aids to avoid overfitting.
As in \citet{simpson2017},
we use an exponential prior for the distance $d(\cdot)$ between the two models,
and transform this to a prior for the desired parameter.
This means that the PC prior always is an exponential distribution on the distance,
while for the parameter in question $\theta$
the distribution varies with parameterization, choice of base model and covariance matrix structures.
The PC prior does in general not have an analytical expression.
The parameter $\theta$ can be a variance or standard deviation \citep{simpson2017}, 
a variance proportion \citep{fuglstad2020,hem2021}, or, for example, a correlation parameter \citep{guo2017}.
In \pkg{makemyprior} we consider 
standard deviations $\sigma$ (and variances $\sigma^2$) 
together with
variance proportions $\omega$.
For a standard deviation the distance is simply $d(\sigma) = \sigma$
\citep{simpson2017}. For a variance proportion parameter the distance will be a function of the covariance matrices of the random effects 
involved in the split, see \citet{fuglstad2020} for details.

\subsubsection{Shrinkage priors for variance proportions}

We construct the joint hierarchical decomposition (HD) prior using 
a bottom-up approach following the prior tree, and
the prior will thus be dependent on prior tree structure.
The distance measure $d(\cdot)$ 
for a variance proportion
depends on the covariance matrices of the effects
of the child nodes
in a split \citep{fuglstad2020}, 
and the covariance matrix of a split node will be a function of the
variance proportions(s) involved in this split.
This means we must condition on the variance proportions 
associated with splits lower in the tree (if any),
and that each prior depends on choices and covariance matrices at 
that and lower levels.
We omit the dependence of tree structure, covariance matrices and
prior choices for other splits in the notation of the PC prior for 
readability. 
Consider a random intercept model 
$y_{i,j} = a_i + \varepsilon_{i,j}$ for $i,j = 1, \dots, 10$,
where $a_i \overset{\text{iid}}{\sim} \mathcal{N}(0, \sigma_{\mathrm{a}}^2)$ is a group
effect and
$\varepsilon_i \overset{\text{iid}}{\sim} \mathcal{N}(0, \sigma_{\varepsilon}^2)$ is
a residual effect.
We define the variance proportion 
$\omega = 
\frac{\sigma_{\mathrm{a}}^2}{\sigma_{\mathrm{a}}^2 + \sigma_{\varepsilon}^2}$.
Then we
denote the
different PC prior distributions as:
\begin{itemize}
	\item $\sigma_{\mathrm{*}} \sim \mathrm{PC}_{\mathrm{0}}(U, \alpha)$, with 
	$\mathrm{Prob}(\sigma_{\mathrm{*}} > U) = \alpha$, 
	and shrinkage towards $\sigma_{\mathrm{*}} = 0$.
	\item $\omega \sim \mathrm{PC}_{\mathrm{0}}(m)$ with 
	$\mathrm{Prob}(\omega > m) = 0.5$ so that $m$ defines the median, and shrinkage towards
	$\omega = 0$, 
	i.e., the base model is a model with only $\bm{\varepsilon}$.
	\item $\omega \sim \mathrm{PC}_{\mathrm{1}}(m)$ with 
	$\mathrm{Prob}(\omega > m) = 0.5$ so that $m$ defines the median, and shrinkage towards
	$\omega = 1$, 
	i.e., the base model is a model with only $\bm{a}$.
	\item $\omega \sim \mathrm{PC}_{\mathrm{M}}(m, c)$ with
	$\mathrm{Prob}(\omega > m) = 0.5$ and
	$\mathrm{Prob}(\mathrm{logit}(1/4) < \mathrm{logit}(\omega) - \mathrm{logit}(m) < \mathrm{logit}(3/4)) = c$
	so that $m$ defines the median, and $c$ says something about how concentrated the distribution is
	around the median. The shrinkage is towards $\omega = m$,
	i.e., the base model is a combination of the effects $\bm{a}$ and $\bm{\varepsilon}$.
\end{itemize}
Note that $\mathrm{PC}_{\mathrm{1}}(m)$ for $\omega$ is equivalent to
$\mathrm{PC}_{\mathrm{0}}(1-m)$ for $1-\omega = \frac{\sigma_{\varepsilon}^2}{\sigma_{\mathrm{a}}^2 + \sigma_{\varepsilon}^2}$.
Since the PC prior is a prior put on the distance between two models, and then transformed to 
the parameter of interest, we do not distinguish the notation 
between the PC prior on a standard deviation
and variance parameter, as it will result in the same prior.

In Figure \ref{fig:pc_prior} we show examples for the four different priors.
Code producing these graphs can be found in Appendix \ref{app:codefig2}.
The shape of the PC prior on a standard deviation, shown in
Figure \ref{fig:pc0_s_prior}, is independent of the hyperparameters and
other hyperparameters will simply give a rescaling of the axes.

Figure \ref{fig:pc0_w_prior} shows a prior where we have shrinkage towards
$\omega = 0$, which corresponds to a model with only residual variance.
In Figure \ref{fig:pc1_prior} we show a prior where with shrinkage
towards $\omega = 1$, giving a model with only group effect.
This base model has a singular covariance matrix, giving a 
distance measure that is infinite.
In such cases we cannot have a median
that is further than $0.25$ from the base model
(see \citet[][Theorem 1]{fuglstad2020} for details).
In this case that means that the median cannot be smaller than $0.75$.

The covariance matrix of the base model is always non-singular for a 
$\mathrm{PC}_{\mathrm{M}}$ prior (shown in Figure \ref{fig:pcm_prior}).
The concentration parameter $c$ in $\mathrm{PC}_{\mathrm{M}}(m, c)$ measures 
how certain we are about the prior median.
A concentration of less than $0.5$ will indicate that a prior with median at
$\omega = 0.5$ is less concentrated around $0.5$ than a uniform distribution
on $[0, 1]$ would be, and we set the lower limit of $c$ to $0.5$.
As this parameter says something about how much of the distribution
mass is in an interval that is smaller than the parameter space 
($0 \leq \omega \leq 1$), the distribution will not change much when $c$
approaches $1$.

\begin{figure}
     \centering
     \begin{subfigure}[t]{0.4\textwidth}
         \centering
         \includegraphics[width=\textwidth]{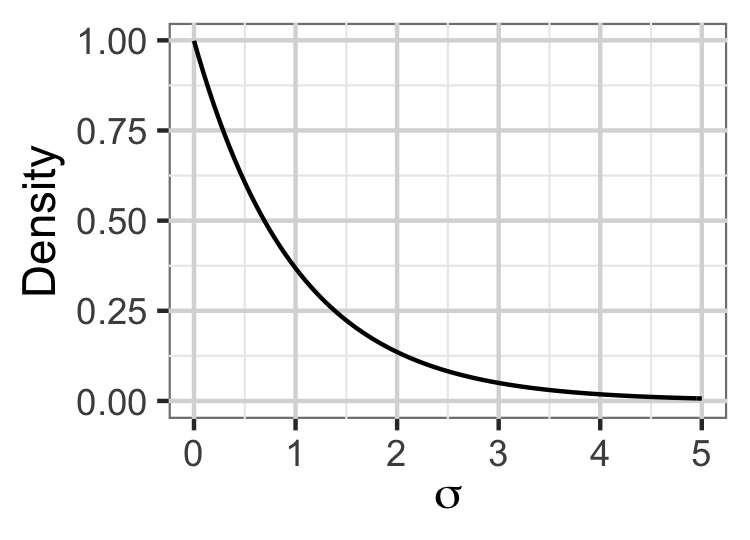}
         \caption{$\sigma \sim \mathrm{PC}_{\mathrm{0}}(3, 0.05)$.}
         \label{fig:pc0_s_prior}
     \end{subfigure}
     \begin{subfigure}[t]{0.4\textwidth}
         \centering
         \includegraphics[width=\textwidth]{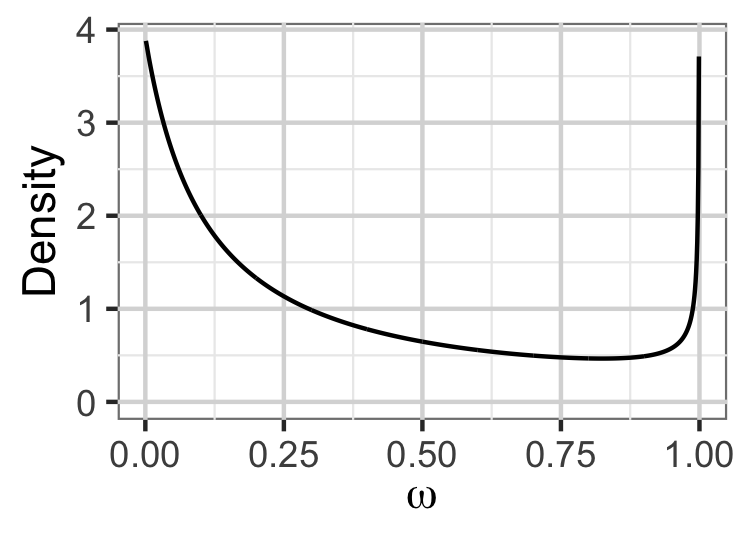}
         \caption{$\omega \sim \mathrm{PC}_{\mathrm{0}}(0.25)$.}
         \label{fig:pc0_w_prior}
     \end{subfigure}
     
     \begin{subfigure}[t]{0.4\textwidth}
         \centering
         \includegraphics[width=\textwidth]{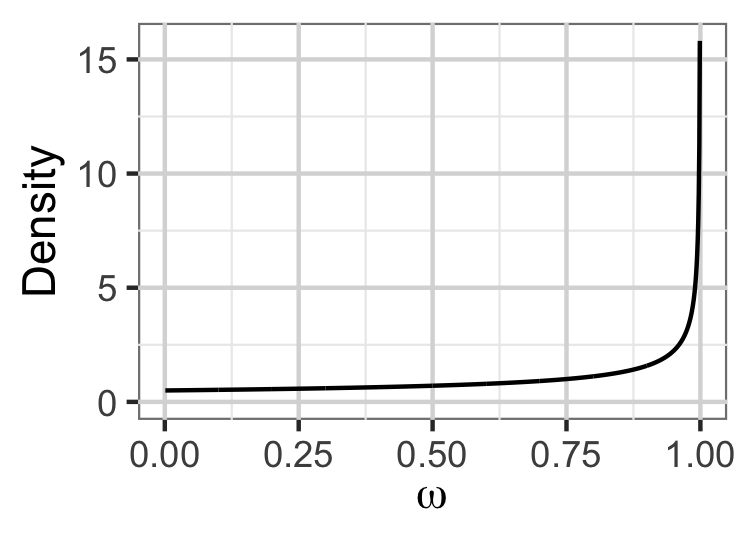}
         \caption{$\omega \sim \mathrm{PC}_{\mathrm{1}}(0.75)$. Note that the base model has a singular covariance matrix.}
         \label{fig:pc1_prior}
     \end{subfigure}
     \begin{subfigure}[t]{0.4\textwidth}
         \centering
         \includegraphics[width=\textwidth]{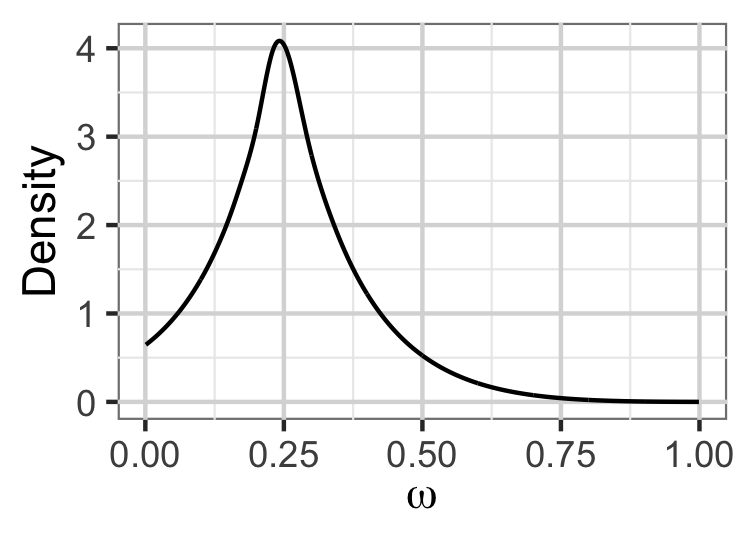}
         \caption{$\omega \sim \mathrm{PC}_{\mathrm{M}}(0.25, 0.85)$.}
         \label{fig:pcm_prior}
     \end{subfigure}
        \caption{Examples of the different PC priors for a random intercept model
        $y_{i,j} = a_i + \varepsilon_{i,j}$ for $i,j = 1, \dots, 10$.
        $\omega = \sigma_{\mathrm{a}}^2/(\sigma_{\mathrm{a}}^2 + \sigma_{\varepsilon}^2)$.}
        \label{fig:pc_prior}
\end{figure}

The difference between the $\mathrm{PC}_{\mathrm{0}}$/$\mathrm{PC}_{\mathrm{1}}$ 
and $\mathrm{PC}_{\mathrm{M}}$ priors is how
certain the user is in the prior knowledge.
In the 
genomic model in Example \ref{exmp:genomic},
a geneticist told us that the 
heritability $h^2$
is around $0.4$.
If the contribution of the additive effect is unclear,
for example
due to a small data sample, we can use
$\mathrm{PC}_{\mathrm{0}}(0.4)$ for $h^2$, which has shrinkage towards
$0$ (only residual effect) and median $0.4$.
If we are certain that 
the additive genetic effect is
contributing to the variation in the observed phenotype, we use a $\mathrm{PC}_{\mathrm{M}}$ prior with median $m = 0.4$, and we
choose the concentration parameter $c$ based on how strongly we believe in this value.
This gives
a prior with shrinkage towards a heritability of $0.4$.

\subsubsection{Ignorance priors for variance proportions}

If we want to express ignorance about the variance distribution in a split,
we use a symmetric Dirichlet prior distribution. 
For a split with $p \geq 2$ children, this prior is given by:
\[
    \pi(\bm{\omega}; p) = \mathrm{Dirichlet}(p) = 
    \frac{\Gamma\left(p\alpha\right)}{\Gamma\left(\alpha\right)^p} 
        \left(\prod_{i=1}^p \omega_i\right)^{\alpha-1},
\]
where $\bm{\omega}$ is 
the vector of variance proportions involved in the split
with $0 < \omega_i < 1$ for $i=1, \ldots, p$ and $\sum_{i=1}^p \omega_i = 1$. Further,
$\Gamma(\cdot)$ denotes the gamma function and $\alpha > 0$ is chosen so 
$\mathrm{P}(\mathrm{logit}(1/4) < \mathrm{logit}(w_i) - \mathrm{logit}(1/p) < \mathrm{logit}(3/4)) = 1/2$ for $i = 1, \dots, p$ (if this is achieved for one $i$, it is by symmetry achieved for all $i$).
This prior assigns equal amount of variance to each model component.
For a split consisting of $p$ nodes we denote this as
$\bm{\omega} \sim \text{Dirichlet}(p)$ for each proportion in the split.
A multi-split is by itself a way of showing ignorance, and all
multi-splits are given this Dirichlet prior.

\subsubsection{Shrinkage priors for multi-splits}

There may be situations where the user wants to assign unequal amounts of variance to the components
in a multi-split. In that case, the user has opinions about the variance decomposition in the split, 
and an ignorant multi-split with a Dirichlet prior is not suitable.
Instead we transform the multi-split to several dual splits and assign 
a PC prior to each of the dual splits.
In Example \ref{exmp:doe}, assume that we want a (20, 30, 50) division of the
total variance
of $\bm{\alpha}$, $\bm{\beta}$ and $\bm{\gamma}$. 
We achieve this by first splitting the variance
50/50 between $\bm{\gamma}$ and $\bm{\alpha}$, $\bm{\beta}$ with a $\mathrm{PC}_{\mathrm{M}}(0.5, c)$
prior, and then dividing the variance of $\bm{\alpha}$ and $\bm{\beta}$
40/60 with a $\mathrm{PC}_{\mathrm{M}}(0.4, c)$ prior for
some suitable value of $c$.
The corresponding tree structure is shown in Figure \ref{fig:theory:trees_c}.
We could also have chosen another effect to split off first, 
but \citet{fuglstad2020} show that
this order does not have much impact on the resulting prior.

\subsubsection{Pre-computing the marginal priors}
 
To ease the
computation of the joint PC prior, we follow \cite{fuglstad2020} and condition on the base models of the splits below in the tree, 
instead of on the parameters themselves. 
In this way we can pre-compute the marginal priors for each split in the tree.
The base model for the Dirichlet distribution is equal variance to each component,
i.e., for a split with $p$ components the base model is $1/p$ of the
variance in the split node to each child node. 
A split
with a Dirichlet prior does not use information from lower levels in the tree.

\section{Software overview}
\label{sec:softwareoverview}

The \proglang{R} package \pkg{makemyprior} currently supports latent Gaussian models where the observation model has a Gaussian likelihood with identity link, a Poisson likelihood with log link or a binomial likelihood with logit link. Supported latent Gaussian models can contain fixed and random effects. Random effects can be independent and identically distributed, i.e. $\mathcal{N}(\bm{0}, \sigma^2 \mathbf{I})$, follow a random walk of first or second order (models \code{rw1} and \code{rw2} in \pkg{INLA}), see definition in \citet[Sections 3.3.1 and 3.4.1, respectively]{gmrfbook}, follow the Besag area model (model \code{besag} in \pkg{INLA}), see definition in \citet[Sections 3.3.2]{gmrfbook}, or be user-defined latent Gaussian models controlled by one variance parameter $\sigma^2$ which follow the form $\mathcal{N}(\bm{0}, \sigma^2 \mathbf{Q}^{-1})$, where $\mathbf{Q}$ is a user-defined symmetric positive definite structure matrix. As mentioned earlier fixed effects are of course supported in the linear predictor, but not integrated into the joint prior framework. By default a zero-mean normal prior with a fixed large variance is assigned. 
For variance parameters of a random effect or a total variance, PC priors, half-Cauchy priors, inverse-gamma priors, or in the case of a total variance parameter 
also Jeffreys' prior is supported. Variance proportion parameters can follow the priors introduced in Section \ref{sec:theory:details}, namely $\mathrm{PC}_{\mathrm{0}}$, $\mathrm{PC}_{\mathrm{1}}$,
$\mathrm{PC}_{\mathrm{M}}$ or a symmetric Dirichlet distribution. With these definitions the package covers a wide range of Bayesian hierarchical models and will be of interest for a range of users. 

We mention two recent applications for which \pkg{makemyprior} might have come in handy. \citet{slater-etal-2021} considered an augmented BYM model with a fixed effect, two Besag area models and an unstructured spatial component. The authors reparameterized the model analogously to \citet{riebler2016} and placed a PC prior on the combined spatial variance and a symmetric Dirichlet prior on the weights which distribute the combined spatial variance to the random effect components. The package \pkg{makemyprior} is directly applicable to this model and allows for an equivalent prior derivation. Importantly, it makes it easily feasible and intuitive to incorporate even more prior knowledge on how the variation shall be distributed to the different spatial model components. \citet{franco-etal-2021} apply the framework of \citet{fuglstad2020} to the specific example of variance-partitioning in spatio-temporal disease mapping models. 
Here, the uncertainty is distributed between temporal effects, spatial effects and space-time interactions. The joint prior the authors propose corresponds to defining an underlying prior tree with two levels:
At the upper level the total variance is first distributed between 
the interaction component, and the space-, time-main-effects,
and then at 
the lower level the uncertainty assigned to the main effects 
is distributed between space and time. This model can be further extended 
by using structured and unstructured main effects for time and space. 
The package \pkg{makemyprior} can handle these models directly, and 
will formulate a joint prior based on the user-specific input. 
The prior will be computed without further involvement of the user 
and visualized for inspection.

Posterior inference can be directly computed through \pkg{INLA} and \pkg{rstan}. Additionally, we provide functions to evaluate the joint prior generated by \pkg{makemyprior} at arbitrary parameter values. 
This is suitable for inclusion of the priors in other \proglang{R} packages or
self-written code for inference. A natural first extension of the functionality of \pkg{makemyprior} might be to support Dirichlet distributions with custom hyperparameter values for the variance proportions, or at least a custom defined beta prior in the case of a dual split.

\section[The makemyprior package]{The \pkg{makemyprior} package}
\label{sec:software}

Throughout this section, the use of the \proglang{R} package 
\pkg{makemyprior} is exemplified by
the following model.
\begin{model}[Example model\label{modExample}]

Consider the hierarchical model for the $n = m \cdot p$ observations $y_{i,j}$, $i = 1, \ldots p$ and $j = 1, \ldots, m$,  given by
\begin{align*}
    y_{i,j}|\eta_{i,j}, \sigma_{\varepsilon}^2 &\sim \mathcal{N}(\eta_{i,j}, \sigma_{\varepsilon}^2), \\
    \eta_{i,j} &= \mu + x_i \beta + a_i + b_j,
\end{align*}
where $\mu$ is an intercept, $x_i$ is a covariate with coefficient $\beta$, 
and $a_1, a_2, \ldots, a_p \overset{\text{iid}}{\sim} \mathcal{N}(0, \sigma_\mathrm{a}^2)$
and $b_1, b_2, \ldots, b_m \overset{\text{iid}}{\sim} \mathcal{N}(0, \sigma_\mathrm{b}^2)$ are random effects.
$\varepsilon_1, \varepsilon_2, \dots, \varepsilon_n
\overset{\text{iid}}{\sim} \mathcal{N}(0, \sigma_{\varepsilon}^2)$ are residuals.

\end{model}
The data for this model could be some measures made for year $i$
and age group $j$ with a time-specific covariate $x_i$.
We do not interpret the meaning of the data in this section.


We include four different tree structures for this model in Table \ref{tab:software:example_structures}. 
Several likelihoods, latent models and prior distributions are available
in \pkg{makemyprior}, which can be listed with the function:
\begin{Code}
makemyprior_models(type = c("prior", "latent", "likelihood"), select = NULL)
\end{Code}

\subsection{Specifying the linear predictor and preparing the data object}
\label{sec:software:modelformula}

First, the linear predictor is specified using a
formula object with a syntax similar to e.g. \fct{lm} from \pkg{stats}
and \fct{inla} from \pkg{INLA}.
Covariates are included directly by name in the formula, and 
\fct{mc} is used to include 
information about each random effect:
\begin{Code}
mc(label, model = "iid", ...)
\end{Code}
The main arguments are
\code{label} (name of the effect) and
\code{model} (the type of latent model effect, i.i.d. is the default).
The other arguments depend on
the choice of latent model, see documentation for details.

For Model \ref{modExample} where both $\bm{a}$ and $\bm{b}$ are i.i.d., the formula is:
\begin{CodeChunk}
\begin{CodeInput}
R> formula <- y ~ x + mc(a) + mc(b)
\end{CodeInput}
\end{CodeChunk}
The intercept $\mu$ is included by default, but can be removed with \code{-1}.
The residual effect for a Gaussian likelihood should not be 
specified in the formula.

The second step is to gather data and create a data object as either a \code{data.frame} or \code{list}
with names corresponding to the elements of the formula. For Model \ref{modExample}, we need the observed response \code{y}, the covariate \code{x}, and indexes for \code{a} and \code{b} (these must be specified as integers).
A simple simulated dataset is:
\begin{CodeInput}
R> p <- 10
R> m <- 10
R> n <- m*p
R> 
R> set.seed(1)
R> data <- list(a = rep(1:p, each = m),
+               b = rep(1:m, times = p),
+               x = runif(n))
R> data$y <- data$x + rnorm(p, 0, 0.5)[data$a] +
+    rnorm(m, 0, 0.3)[data$b] + rnorm(n, 0, 1)
\end{CodeInput}
We recommend the use of short names for the input data because these
names need to be used to refer to model components
in later steps. An example is 
\code{"rain"} instead of \code{"rainfall\_august\_2020"}.
Note that the observations $y_{i,j}$ are not used to make the prior.

\subsection{Exploring and selecting the prior graphically}

\begin{table}[t!]
\centering
\begin{tabular}{lcl}
	& \textbf{Tree structure} 
	& \textbf{Text string} \\ \hline \\[-8pt]
	\rotatebox[origin=c]{90}{\textbf{ Prior 1}} & \begin{tabular}{@{}c@{}} \tikz \pic{software_tree3}; \end{tabular} 
	&
	\begin{tabular}{@{}l@{}} \code{"(a); (b); (eps)"} \end{tabular} \\ \hline \\[-8pt]
	\rotatebox[origin=c]{90}{\textbf{Prior 2}} & \begin{tabular}{@{}c@{}} \tikz \pic{software_tree4}; \end{tabular} 
		&
	\begin{tabular}{@{}l@{}} \code{"s1 = (a, b); (eps)"} \end{tabular} \\ \hline \\[-8pt]
	\rotatebox[origin=c]{90}{\textbf{Prior 3}} & \begin{tabular}{@{}c@{}} \tikz \pic{software_tree1}; \end{tabular} 
	&
	\begin{tabular}{@{}l@{}} \code{"s1 = (a, b, eps)"} \end{tabular} \\ \hline \\[-8pt]
	\rotatebox[origin=c]{90}{\textbf{Prior 4}} & \begin{tabular}{@{}c@{}} \tikz \pic{software_tree2}; \end{tabular} 
	&
	\begin{tabular}{@{}l@{}} \code{"s1 = (a, b); s2 = (s1, eps)"} \end{tabular} \\ \hline \\[-8pt]
\end{tabular}
\caption{
Four tree structures for 
Model \ref{modExample} with 
text strings specifying them. The names 
\code{s1} and \code{s2}  on the splits (variance proportions) are chosen by the user in the initial specification of the prior, and are used in a
nested formulation to specify the prior tree structure.
}
\label{tab:software:example_structures}
\end{table}

We provide a graphical user interface (GUI) where
the user can construct the prior in an interactive way.
The GUI is implemented as a
\pkg{shiny} app \citep{shinymanual}
running locally. 
It allows the user to first define the
tree structure, and then be guided sequentially through
the steps of selecting priors for each split, singleton and top node. 
The \pkg{shiny} apps are useful to, for example, 
display and investigate results from analyses.
For example, 
the user can 
customize graphs and tables in a simple way.
Such
apps are used by \citet{depaoli2020} to show why prior sensitivity analysis is important,
and by \citet{smid2020} to let users explore the impact of prior distributions in
inference.
However, to the extent of our knowledge, as of today there are no packages or apps that allows the user to
use a \pkg{shiny} app (or similar) to specify priors for custom models and data and directly
carry out inference. 

The first step is to initialize a prior object. 
This is done with the function
\fct{make\_prior}: 
%
\begin{Code}
make_prior(formula, data, family = "gaussian", prior = list(), 
           intercept_prior = c(), covariate_prior = list())
\end{Code}
\code{formula} and \code{data} are the objects created
in Section \ref{sec:software:modelformula}, and 
\code{family} is the likelihood (\code{"gaussian"} is the default, 
\code{"binomial"} and \code{"poisson"} are also available). 
The \code{prior} argument is not relevant when the
GUI is used, and its 
description is deferred to Section \ref{sec:software:make_prior}. 
\code{intercept\_prior} and \code{covariate\_prior} specifies the mean
and standard deviation for the Gaussian
priors on the intercept
and covariate coefficients.
The default in \pkg{makemyprior} is $\mathcal{N}(0, \sigma = 1000)$ for both, and the
coefficient priors are specified as a named list with names corresponding to the covariate names.

For Model \ref{modExample}, we can create
a prior object with the following command:
\begin{CodeChunk}
\begin{CodeInput}
R> prior <- make_prior(formula, data, family = "gaussian", 
+                      intercept_prior = c(0, 1000), 
+                      covariate_prior = list(x = c(0, 100)))
\end{CodeInput}
\begin{CodeOutput}
Warning message:
Did not find a tree, using default tree structure instead.
\end{CodeOutput}
\end{CodeChunk}
This gives a prior with a single tree with one split as shown in Prior 3 in Table \ref{tab:software:example_structures}, which is the default setting.
A warning makes the user aware that the default prior is chosen.
For Gaussian likelihoods,
Jeffreys' prior is by default set for the total variance. 
Let $\sigma_{\mathrm{a+b} + \varepsilon}^2$ denote total variance, 
and 
$\boldsymbol{\omega} = \big(\omega_{\frac{\mathrm{a}}{\mathrm{a+b} + \varepsilon}}, \omega_{\frac{\mathrm{b}}{\mathrm{a+b} + \varepsilon}}, 1-\omega_{\frac{\mathrm{a}}{\mathrm{a+b} + \varepsilon}}-\omega_{\frac{\mathrm{b}}{\mathrm{a+b} + \varepsilon}} \big)$ describe the attribution of variance
to the three different sources, then the initial
choice of priors is:
\begin{equation}
    \bm{\omega} \sim \mathrm{Dirichlet}(3) \quad \text{and}\quad \sigma_{\mathrm{a+b} + \varepsilon}^2 \sim \mathrm{Jeffreys'}.
    \label{eq:software:default_prior}
\end{equation}
The intercept has a $\mathcal{N}(0, \sigma = 1000)$ prior, and the covariate $\beta$
a $\mathcal{N}(0, \sigma = 100)$ prior.

\begin{figure}[t]
    \centering
    \includegraphics[width = 1\textwidth]{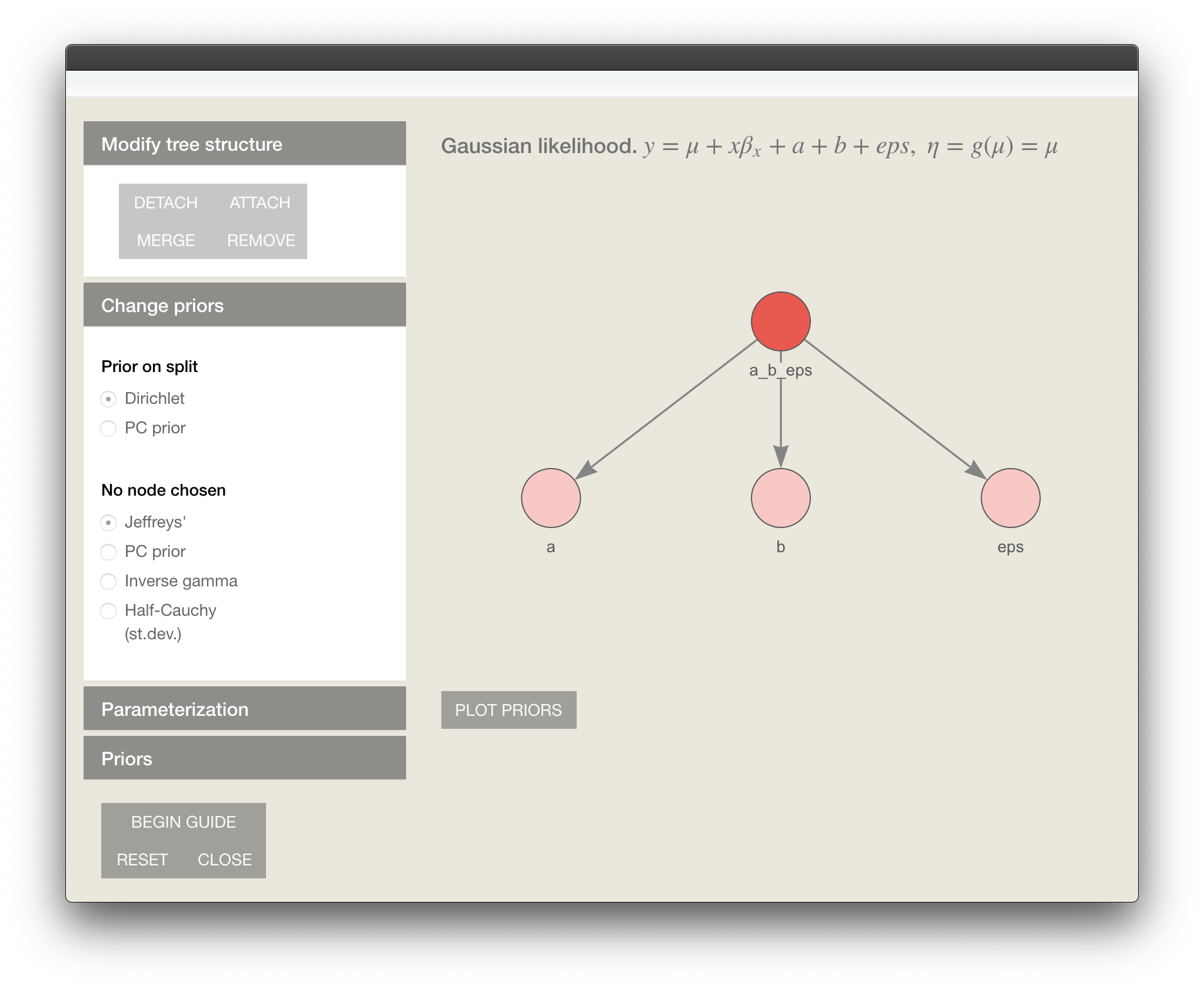}
    \caption{Screenshot of the GUI in \pkg{makemyprior} with the 
    default prior
    (Equation \ref{eq:software:default_prior}).}
    \label{fig:software:default_app}
\end{figure}

The function \fct{makemyprior\_gui} allows the user to
select the desired prior tree structure and choose prior distributions 
interactively: 
\begin{Code}
makemyprior_gui(prior, guide = FALSE, no_pc = FALSE)
\end{Code}
This function takes the arguments \code{prior}, created with
\fct{make\_prior} earlier,
\code{guide} which specifies whether or not the guide should automatically start
(the guide can be started at any time), and
\code{no\_pc}. 
The PC prior is computed using the covariance matrix structure of the model components, which
may be slow for large models. 
For a better user experience,
the user can turn off the computation
of the PC prior in the GUI using \code{no\_pc = TRUE}. 
The prior will be computed upon closing, and this will only affect the plotting of the prior
in the GUI.

For Model \ref{modExample}, we start the GUI by 
running:
\begin{CodeChunk}
\begin{CodeInput}
R> new_prior <- makemyprior_gui(prior)
\end{CodeInput}
\end{CodeChunk}
This saves the new prior chosen
to the variable \code{new\_prior} when we close the GUI. 
Figure 
\ref{fig:software:default_app} shows a screenshot of the GUI for Model \ref{modExample}
for $m = p = 10$ and the default prior. 
By selecting nodes and using the buttons located in the panels to the left,
the user can create the desired tree structure 
and choose the desired priors for this tree.

\begin{figure}
    \begin{subfigure}[t]{0.32\textwidth}
        \includegraphics[width = \textwidth]{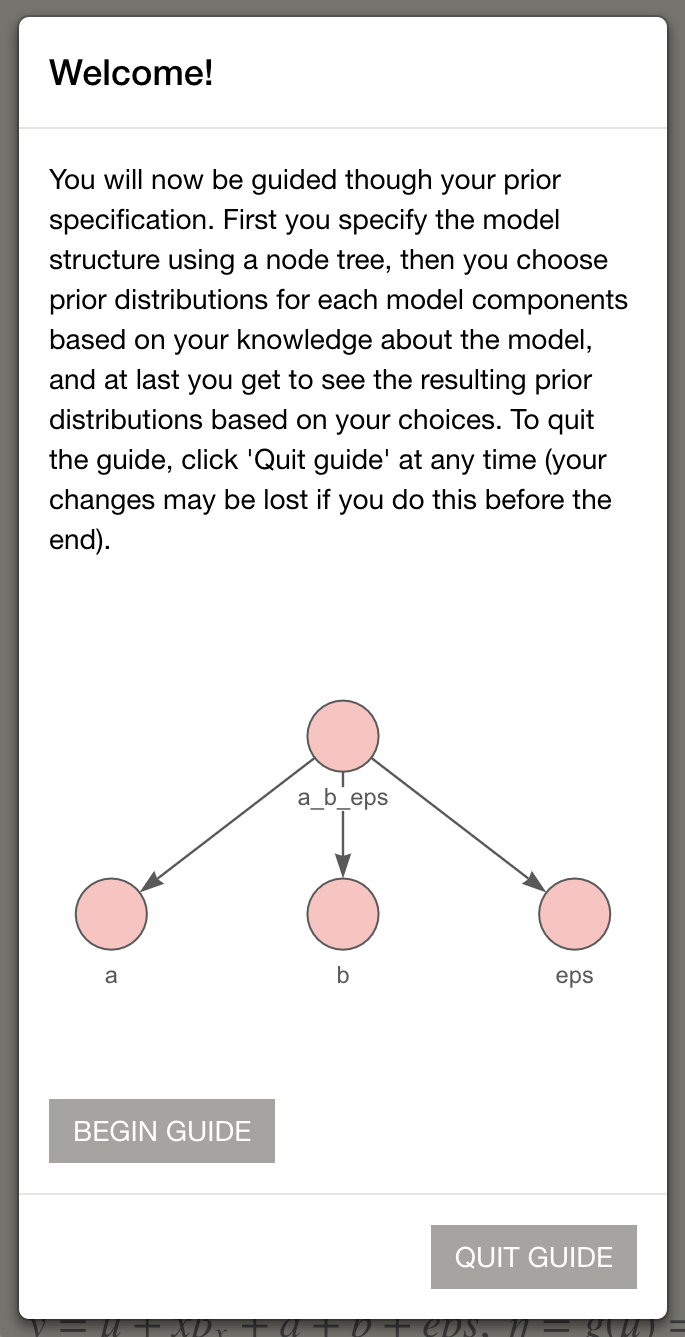}
        \caption{Initial window.}
        \label{fig:software:guide1}
    \end{subfigure}
    \begin{subfigure}[t]{0.32\textwidth}
        \includegraphics[width = \textwidth]{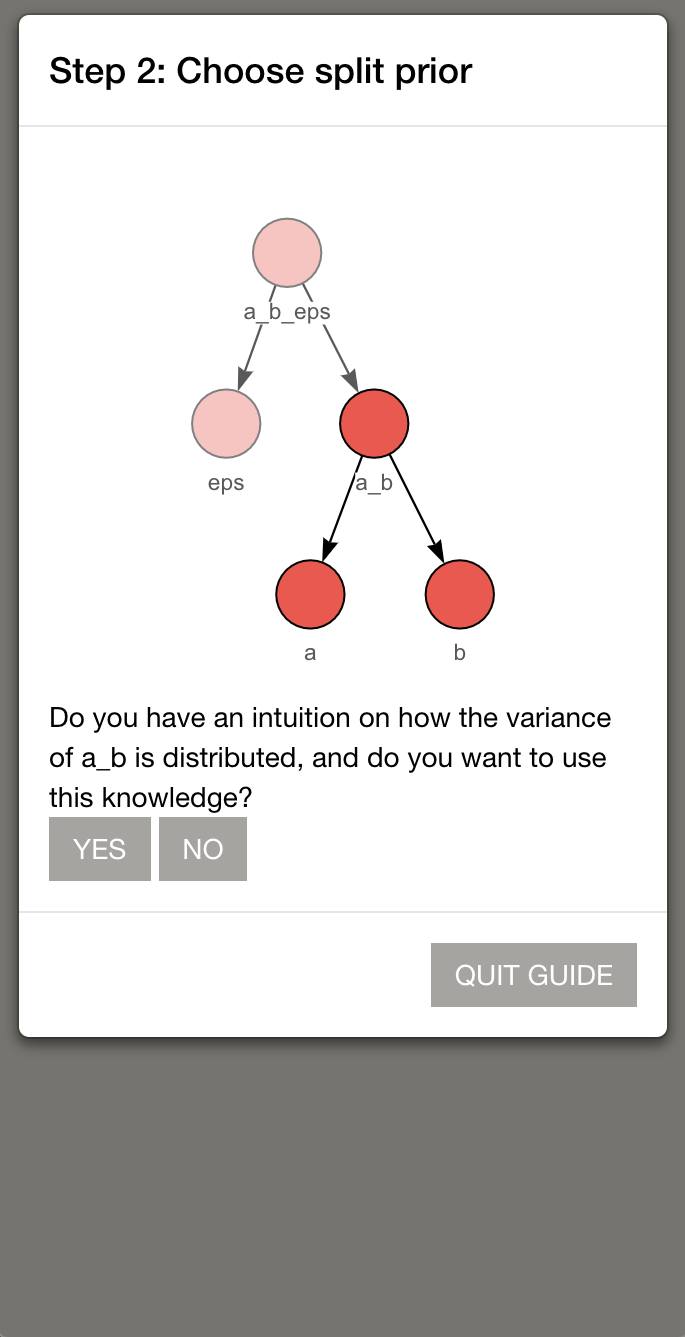}
        \caption{Question for prior elicitation.}
        \label{fig:software:guide2}
    \end{subfigure}
    \begin{subfigure}[t]{0.32\textwidth}
        \includegraphics[width = \textwidth]{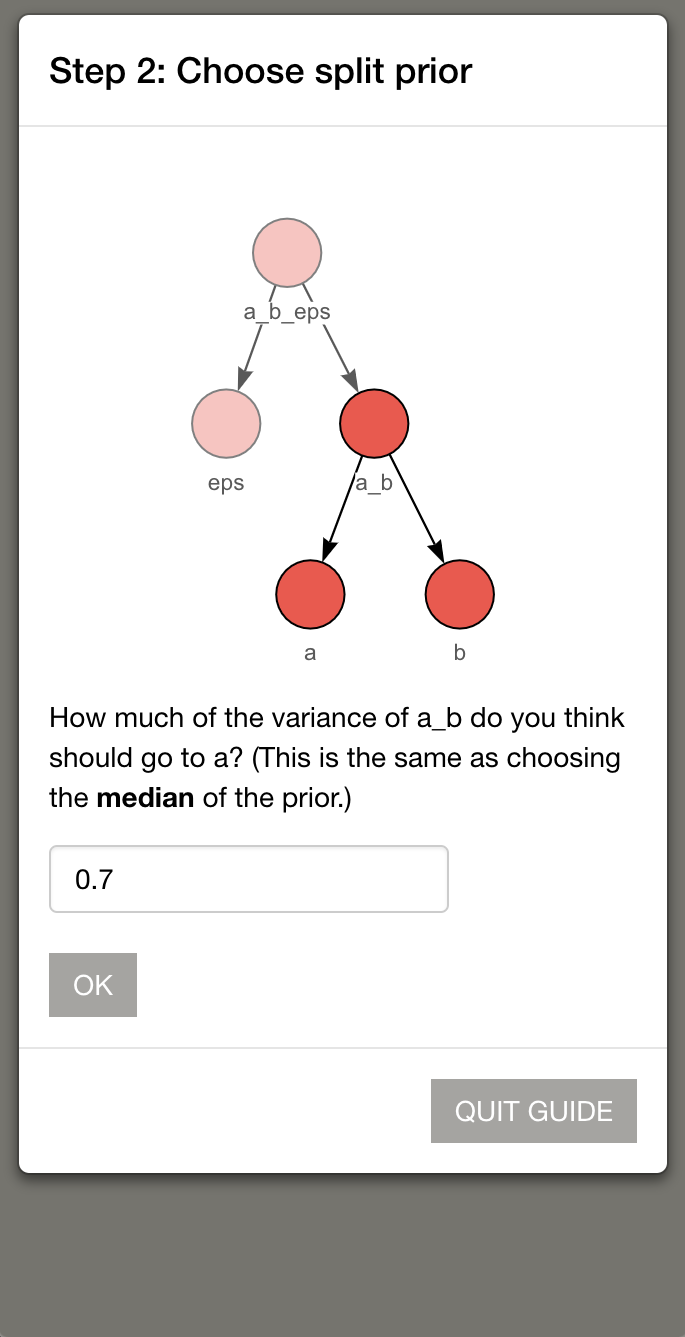}
        \caption{Question for prior elicitation.}
        \label{fig:software:guide3}
    \end{subfigure}
    \caption{Screenshots from the guide in the \pkg{makemyprior} GUI.}
\end{figure}

\begin{figure}[t]
    \centering
    \includegraphics[width = 1\textwidth]{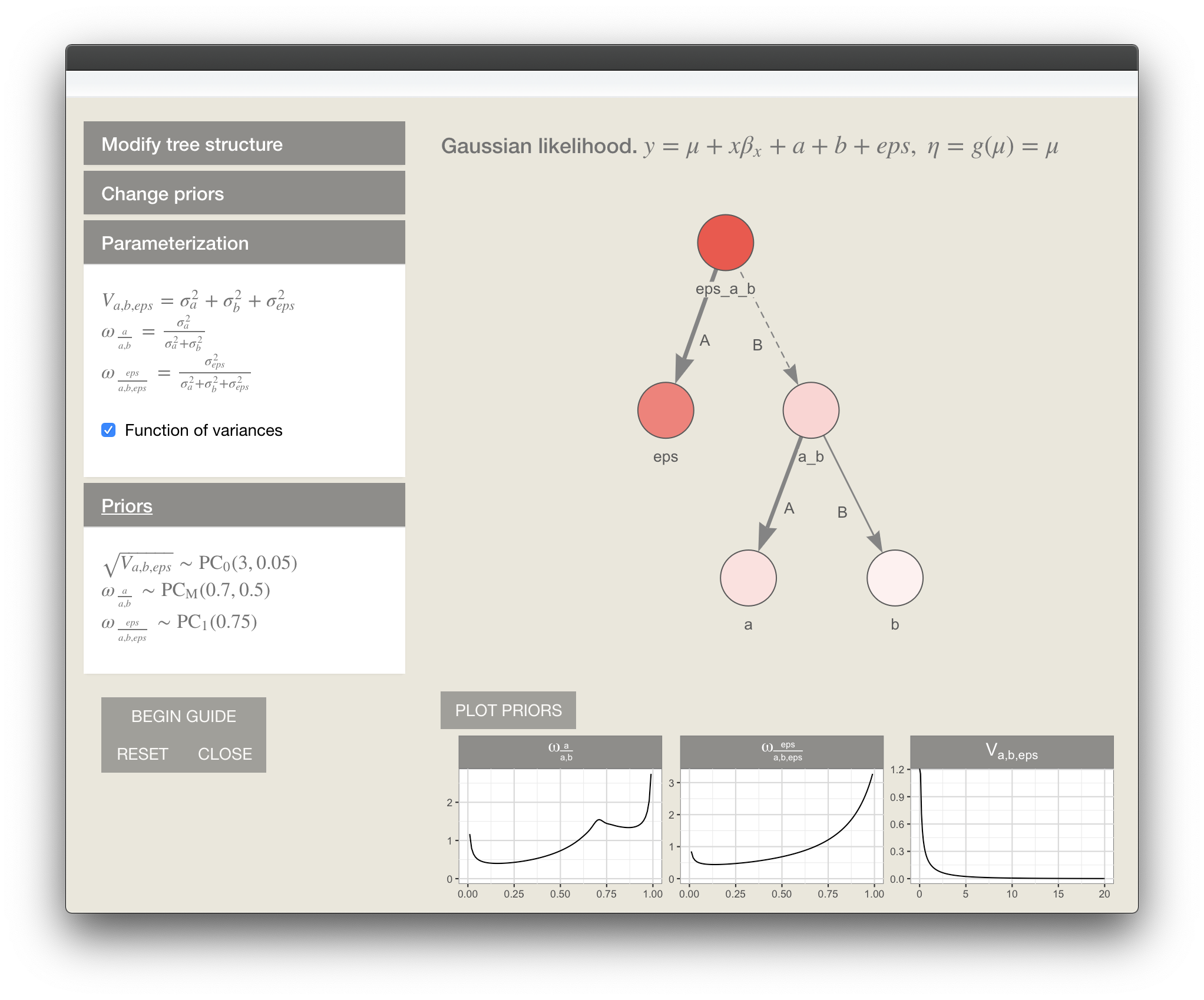}
    \caption{Screenshot of the GUI in \pkg{makemyprior} when the prior 
    in 
    Equation \ref{eq:software:examplemodel_prior} is
    chosen.}
    \label{fig:software:modified_app}
\end{figure}

The user can click "Begin guide" to the lower left
to be guided through the prior construction.
Figure \ref{fig:software:guide1} shows the initial window
of the guide, which consists of two main steps: 
First, the user is helped to
create the desired tree structure.
Note that every time the tree structure is modified, both when 
using the guide and otherwise, all splits are 
set to have the default
Dirichlet prior.
After deciding on a tree structure,
the user will be taken through the prior tree structure in a 
step-wise manner. 
For each
split, top node and singleton they will be asked simple questions about
the existing prior knowledge about the corresponding parameter.
This eases the process of using the knowledge to create the prior, as
the user does not need to think in terms of prior distributions directly,
but only needs to answer questions such as shown in Figure \ref{fig:software:guide2} and \ref{fig:software:guide3}.
After modifying the prior to have the following distributions:
\begin{equation}
    \omega_{\frac{\mathrm{a}}{\mathrm{a+b}}} \sim \mathrm{PC}_{\mathrm{M}}(0.7, 0.5),\,  \omega_{\frac{\mathrm{a+b}}{\mathrm{a+b} + \varepsilon}} \sim \mathrm{PC}_{\mathrm{0}}(0.25),\,\text{and}\, \sigma_{\mathrm{a+b} + \varepsilon} \sim \mathrm{PC}_{\mathrm{0}}(3, 0.05),
    \label{eq:software:examplemodel_prior}
\end{equation}
the GUI looks like the screenshot in Figure \ref{fig:software:modified_app}.
The chosen prior distributions and the 
connection between the parameterization and model variances are easily 
seen to the left.

The GUI is intuitive and contains a thorough description of the options,
and we do not explain the features
in detail here. 
We instead recommend using the guide in the GUI
to get familiar with the package.
A summary of the prior object can be printed with:
\begin{CodeChunk}
\begin{CodeInput}
R> summary(new_prior)
\end{CodeInput}
\begin{CodeOutput}
Tree structure: a_b = (a,b); eps_a_b = (eps,a_b)

Weight priors:
	w[a/a_b] ~ PCM(0.7, 0.5)
	w[eps/eps_a_b] ~ PC1(0.75)
Total variance priors:
	sqrt(V)[eps_a_b] ~ PC0(3, 0.05)

Covariate priors: intercept ~ N(0, 1000^2), x ~ N(0, 100^2)

y ~ x + mc(a) + mc(b)
\end{CodeOutput}
\end{CodeChunk}

\subsection{Selecting the prior non-graphically}
\label{sec:software:make_prior}

If the prior has been constructed with the
GUI, this section can be skipped.
Here we describe an alternative way to specify 
the prior without the use of the GUI.
This is
done with same function we use to make the default prior, \fct{make\_prior}, by specifying the 
\code{prior} argument. \code{prior} is a named \code{list} with the following arguments:
\begin{description}
    \item [\code{tree}] The tree structure as a string. A split is specified as \code{s1 = (a, b)}, where \code{s1} represents a split node and can be any name except
    names of the input data in \code{data} and the reserved \code{eps}, which is used for residuals for a 
    Gaussian likelihood. Short names
    are recommended. Note that these split names are just used in the initial specification.
    The child nodes for each split are included in parentheses separated by commas, and each split is separated by semicolons. 
    Singletons are included as \code{(a)}.
    Examples of strings for different tree structures for 
    Model \ref{modExample} are
    shown in Table \ref{tab:software:example_structures}.
    \item [\code{V}] A named list with information on the priors on each top node and singleton,
    i.e., all variances. Options are 
    \code{"pc"}, \code{"jeffreys"}, \code{"invgam"}, and \code{"hc"} (Half-Cauchy).
    The names in the list are the top node and singleton names from the \code{tree} argument.
    \item [\code{w}] A named list with information on the priors on each split,
    i.e., all variance proportions. 
    The names in the list are the split node names from the \code{tree} argument.
    Options are \code{"pc0"}, \code{"pc1"}, \code{"pcM"} and \code{"dirichlet"}.
\end{description}
\code{V} and \code{w}
must have the following 
structure for each element in the list: 
\code{list(prior = prior\_name, param = parameter\_vector)}, except for 
Jeffreys' and Dirichlet priors, where \code{param} is not specified 
(as the distributions do not have hyperparameters).
See Section \ref{sec:examples} for how to specify priors in
specific examples,
and \fct{makemyprior\_models} for more 
details about the different prior distributions.

The prior in Equation \ref{eq:software:examplemodel_prior} for Model \ref{modExample} can be specified as:
\begin{CodeChunk}
\begin{CodeInput}
R> prior <- make_prior(
+    formula, data,
+    prior = list(
+      tree = "s1 = (a, b); s2 = (s1, eps)",
+      w = list(s1 = list(prior = "pcM", param = c(0.7, 0.5)),
+               s2 = list(prior = "pc0", param = 0.25)),
+      V = list(s2 = list(prior = "pc0", param = c(3, 0.05)))
+    ),
+    covariate_prior = list(x = c(0, 100)))
R> prior
\end{CodeInput}
\begin{CodeOutput}
Model: y ~ x + mc(a) + mc(b)
Tree structure: a_b = (a,b); eps_a_b = (eps,a_b)

Weight priors:
	w[a/a_b] ~ PCM(0.7, 0.5)
	w[eps/eps_a_b] ~ PC1(0.75)
Total variance priors:
	sqrt(V)[eps_a_b] ~ PC0(3, 0.05)

Covariate priors: intercept ~ N(0, 1000^2), x ~ N(0, 100^2)
\end{CodeOutput}
\end{CodeChunk}
The names \code{s1} and \code{s2} are chosen by the user in
the initial specification of the prior, and can be any names that are not
used for the data or \code{eps} (reserved for residuals).
Note that \code{s1} and \code{s2}
are
only used as a link between the splits and priors, and
have been automatically changed to \code{a\_b} and
\code{eps\_a\_b} 
by \fct{make\_prior}.
The order we list the children for each split node in the \code{tree} argument 
decides 
which way we have shrinkage
with the PC priors.
For \code{s1 = (a, b)}, $\mathrm{PC}_{\mathrm{0}}(m)$ 
shrinks effect $\bm{a}$ 
($\omega_{\frac{\mathrm{a}}{\mathrm{a+b}}} = 0$ as base model), 
$\mathrm{PC}_{\mathrm{1}}(m)$ 
shrinks effect $\bm{b}$
and $\mathrm{PC}_{\mathrm{M}}(m, c)$ gives shrinkage towards $m \bm{a} + (1-m) \bm{b}$.
All three has median at $\omega_{\frac{\mathrm{a}}{\mathrm{a+b}}} = m$. 
Note that $\mathrm{PC}_{\mathrm{0}}(m)$ on $\omega_{\frac{\mathrm{a}}{\mathrm{a+b}}}$ is equivalent to
$\mathrm{PC}_{\mathrm{1}}(1-m)$ on $\omega_{\frac{\mathrm{b}}{\mathrm{a+b}}}$.

All top nodes, singletons and split nodes without a specified prior will get the default prior.
The default settings in \pkg{makemyprior} are chosen based on
the findings of \cite{fuglstad2020} to ensure robust inference:
\begin{itemize}
    \item If no prior is specified (neither tree structure nor priors), the prior will be
    a joint prior where all latent components (including
    a possible residual effect) get an equal amount of the total variance in the prior through
    the symmetric Dirichlet prior, and the default total variance prior.
    \item The default prior on the total variance (top node) varies with likelihood:
    \begin{itemize}
        \item Jeffreys' prior for Gaussian likelihood for a tree structure with one tree, $\mathrm{PC}_{\mathrm{0}}(3, 0.05)$ otherwise.
        \item $\mathrm{PC}_{\mathrm{0}}(1.6, 0.05)$ for binomial likelihood.
        \item $\mathrm{PC}_{\mathrm{0}}(1.6, 0.05)$ for Poisson likelihood.
    \end{itemize}
    \item The default prior on an individual variance (singleton) varies with likelihood: 
    \begin{itemize}
        \item $\mathrm{PC}_{\mathrm{0}}(3, 0.05)$ for Gaussian likelihood.
        \item $\mathrm{PC}_{\mathrm{0}}(1.6, 0.05)$ for binomial likelihood.
        \item $\mathrm{PC}_{\mathrm{0}}(1.6, 0.05)$ for Poisson likelihood.
    \end{itemize}
    \item The default prior on a variance proportion (split node) is a Dirichlet prior assigning equal amount
    of variance to each of the model components involved in the split.
\end{itemize}
The reasoning behind these choices are as follows.
For the variance proportions, there is no way of knowing what behaviour 
is desired in general. Therefore we use the ignorant Dirichlet prior as
the default.
A standard Gaussian distribution with mean $0$ and variance $1$ will have close 
to all the density mass
between $-3$ and $3$, and $\mathrm{PC}_{\mathrm{0}}(3, 0.05)$ is a vague
prior for the standard deviation in such a distribution.
We use Jeffreys' prior when applicable (see Section \ref{sec:theory:simple}).
To choose the default variance (standard deviation) 
prior for other likelihoods, we
follow the idea of \citet{fong2010} and use a 
credible interval on a suitable scale.
The default prior for all variance 
parameters (both for top nodes and singletons) in \pkg{makemyprior} for
binomial and Poisson likelihoods is a PC prior with a 
95\% credible interval between $0.2$ and $5$
for the multiplicative effect on the odds ratio and risk, respectively.
This is 
obtained with a $\mathrm{PC}_{\mathrm{0}}(1.6, 0.05)$ prior.
We want to emphasize that before selecting the default prior, both when
using \pkg{makemyprior} and otherwise,
you should stop and think about whether or not it is suitable for your model 
and data.





	


\subsection{Performing inference}
\label{sec:software:inference}

We include functions 
for inference that are compatible with the prior object obtained from \fct{make\_prior} (and \fct{makemyprior\_gui}).
Both 
\proglang{Stan} \citep{carpenter2017} through \pkg{rstan} \citep{rstan2020} and
Integrated Nested Laplace Approximations \citep[INLA,][]{rue2009} through \pkg{INLA} (see \url{www.r-inla.org})
can be used
for the inference.

\proglang{Stan} is a probabilistic programming language, where Hamiltonian Monte Carlo
(HMC)
is used to sample from the posterior distribution
\citep{carpenter2017}. 
\proglang{Stan} implements HMC using the No U-Turn Sampler \citep[NUTS,][]{hoffman2014}.
NUTS reduces the need for tuning of the sampler, making it easy to use as no
manual settings are needed for the algorithm to run, and the user only needs to
provide the joint prior and likelihood model, implemented in a programming
language similar to \proglang{C++}.
We provide \proglang{Stan} code for fitting latent Gaussian
models with certain
likelihoods and latent effects. The internal parameterization
in the provided \proglang{Stan} code is log-variance, and
the prior is transformed from the parameterization given by the prior tree
structure
to log-variances.

INLA is a non-sampling based method for doing fast and efficient 
Bayesian inference on latent Gaussian models \citep{rue2009}, utilizing
Gaussian Markov random fields (GMRFs) with sparse precision matrices,
which gives computational benefits through the Markov property.
The INLA method approximates the posteriors by a mixture of Gaussian distributions
and applies a skewness correction to the marginals \citep{rue2017}.
It is easy and straight-forward to use for inference, and can fit models
with a broad range of latent effects. The internal parameterization
of the model parameters in INLA is log-precision, and in the same
way as for the provided \proglang{Stan} code the parameterization following
the prior tree is transformed to fit INLA.

Some common latent models are included in the code for the package:
i.i.d. (\code{"iid"}), random walk of first (\code{"rw1"}) and second
(\code{"rw2"}) order, Besag (\code{"besag"}), and effects with structured
covariance matrices (\code{"generic0"}). 
The likelihood family was specified in \fct{make\_prior}.
In Section \ref{sec:examples} we show how the inference can be performed.
Here we describe the functions that can be used for inference.

\subsubsection[Inference with Stan]{Inference with \proglang{Stan}}

\proglang{Stan} is a flexible tool for inference, however, it requires the user to
write custom code for the model that is to be fitted. 
\pkg{makemyprior} contains 
pre-written \proglang{Stan}-code that can be used to do inference on latent Gaussian models with
a selection of latent models.
We recommend to compile the \proglang{Stan}-code before doing inference with \proglang{Stan}.
This can be done with the following function:
\begin{Code}
compile_stan(save = FALSE, permanent = FALSE, path = NULL)
\end{Code}
where \code{save} indicates whether or not to save the compiled object 
(must be set this to \code{TRUE} to avoid recompiling the code
every time inference is performed), and
\code{permanent} is set to \code{FALSE} if the compiled model should be saved temporary
in \fct{tempdir} for the current \proglang{R} session, or permanently in the
package directory.
\code{path} is only necessary if the
default location for saving the compiled object is not possible to use 
(see \code{?compile\_stan} for details).
For inference with \proglang{Stan} we use the following function:
\begin{Code}
inference_stan(prior_obj, use_likelihood = TRUE, print_prior = TRUE,
               path = NULL, ...)
\end{Code}
The first argument is the prior object from \fct{make\_prior}
or \fct{makemyprior\_gui}. 
The user can specify whether to include
the likelihood (\code{use\_likelihood = TRUE}) or not 
(\code{FALSE}).
In 
the latter case we sample from the 
prior distribution.
\code{print\_prior} (\code{TRUE} by default) prints details about the chosen prior.
\code{path} is a file path that can be specified if the \code{path} argument were
used in \fct{compile\_stan}, and if left empty, \fct{inference\_stan} looks for the
compiled \proglang{Stan} code in \code{tempdir()}, then the package directory, and if no
compiled code is found the function re-compiles the \proglang{Stan} code.
Additional arguments that is sent directly to the \pkg{rstan} function \fct{sampling} 
can be specified for the inference. Useful arguments include
\code{iter} (total number of iterations for each chain),
\code{warmup} (number of iterations for the warm-up),
\code{chains} (number of chains),
\code{seed} (for reproducibility), and
\code{control} (for specifying algorithm tuning parameters).

The internal parameterization in the \proglang{Stan}-code included in the package is log-variance, however, since
\proglang{Stan} works with samples we can look at any parameterization we want by transforming the log-variances.
For using other latent models or more complex models than the ones provided in the
included \proglang{Stan}-code (see above), 
the user must write customized \proglang{Stan}-code,
see
Section \ref{sec:software:customstan}.

\subsubsection{Inference with INLA}


For inference with INLA we use the following function:
\begin{Code}
inference_inla(prior_obj, use_likelihood = TRUE, print_prior = TRUE, ...)
\end{Code}
The first three arguments are the same as in \fct{inference\_stan}.
Additional arguments can be fed to the \pkg{INLA} function \fct{inla}.
Useful arguments include \code{Ntrials} for the binomial likelihood, used to
specify the amount of trials, where the response is the number of successes.


\subsection{Visualizing priors and posteriors}

We offer several functions to visualize the prior and posterior distributions.
The prior distributions for the random effects on the tree structure 
parameterization can be
plotted with \code{plot\_prior(obj)} which take
an object from \fct{make\_prior}, \fct{makemyprior\_gui}, 
\fct{inference\_stan} or \fct{inference\_inla} as input.
The posterior distributions can be displayed with
\begin{Code}
plot_posterior_variance(obj)
plot_posterior_stdev(obj)
plot_posterior_precision(obj)
\end{Code}
\code{obj} is an object from
\fct{inference\_stan} or \fct{inference\_inla}.

The posterior distributions of random effects
from inference with \proglang{Stan} can be plotted with:
\begin{Code}
plot_posterior_stan(
  obj, 
  param = c("prior", "variance", "stdev", "precision"), 
  prior = FALSE
)
\end{Code}
Here, \code{obj} is an object from \fct{inference\_stan},
\code{param} specifies which parameterization the plots should have where
\code{param = "prior"} gives the posterior on the same parameterization as the prior.
\code{prior} indicates whether or not to plot the prior together with the posterior
for \code{param = "prior"}. The total variance prior will only be plotted
if it is not Jeffreys' prior.
Fixed effect posteriors can be plotted with
\code{plot\_posterior\_fixed(obj)}.
More details about visualization can be found with
\code{?makemyprior\_plotting}.


\subsection[More complex models in Stan]{More complex models in \proglang{Stan}}
\label{sec:software:customstan}

Latent models may have parameters that are not variances, such as correlations. These
non-variance parameters
are handled independently 
in the HD prior \citep{fuglstad2020}.
The \proglang{Stan} code included in \pkg{makemyprior} is
applicable for certain commonly used latent models and likelihoods 
(see Section \ref{sec:software:inference}).
We provide 
a ``skeleton'' code and
a description on how the user can write
custom Stan-code and include the joint prior created with \fct{make\_prior}.
This can be accessed with:
\begin{Code}
create_stan_file(location = "")
\end{Code}
\code{location} is a string to a path where a folder with necessary files
will be stored. 
The user can edit the code and include custom
latent components etc. We do not include details on this, as it will be highly model
specific and is merely an offer to the users who want to apply the HD prior in 
more advanced models.

\subsection{Evaluating the joint prior}
The following functions allow users to construct the joint prior through \pkg{makemyprior}, and evaluate the priors in their own inference code.
With the function 
\fct{eval\_joint\_prior}, the joint HD prior created with \fct{make\_prior}
can be evaluated on log-variance scale:
\begin{Code}
eval_joint_prior(theta, prior_data)
\end{Code}
where \code{theta} is a vector of log-variances and \code{prior\_data} is a condensed prior 
object created with \code{make\_eval\_prior\_data}. The marginal prior distributions for
PC priors on variance proportions can be evaluated using:
\begin{Code}
eval_pc_prior(x, obj, param, logitscale = FALSE)
\end{Code}
where \code{x} is value(s) to evaluate the prior in, \code{obj} is an object
from \fct{make\_prior}, \code{param} is a string indicating which variance proportion we
want to evaluate, and \code{logitscale} indicates whether the input \code{x} is on
logit-scale (\code{TRUE}) or not.

\section[Using makemyprior: Examples]{Using \pkg{makemyprior}: Examples}
\label{sec:examples}

In this section we provide three examples where we use the \pkg{makemyprior} package
to construct priors and run inference. Two examples are with Gaussian responses, and one is with
Binomial responses.
We have used \proglang{Stan} for the inference (with \fct{inference\_stan}), 
but the procedure is the same for
inference with INLA (using \fct{inference\_inla} instead).

\subsection{Gaussian responses}

\subsubsection{Genomic selection in wheat breeding}

This is an extended version of the model in Example \ref{exmp:genomic}. 
In addition to the additive genetic effect $\bm{a}$, we now also include
two nonadditive effects: Dominance $\bm{d}$ and additive-by-additive epistasis
$\bm{x}$.
This example is taken from \citet{hem2021}. 
The response $y_i$ is grain yield for individual $i$. We utilize the expert knowledge elicited from experts in the field, and
create a prior distribution reflecting this knowledge. 
We model the response as:
\begin{equation}
    y_i = \mu + a_i + d_i + x_i + \varepsilon_i, \ i = 1, \dots, 100,
\end{equation}
where $\mu$ is an intercept with default $\mathcal{N}(0, 1000^2)$ prior
and $\varepsilon_i$ is the residual effect, 
representing environmental noise. Further, $a_i$, $d_i$ and $x_i$ are 
additive, dominance and epistasis (additive-by-additive epistasis) effects, respectively. 
These three add up to the genetic effect $g_i = a_i + d_i + x_i$.
We assume that
$\bm{a} = (a_1, \dots, a_{100}) \sim \mathcal{N}_{100}(\bm{0}, \sigma_{\mathrm{a}}^2 \bm{A})$,
$\bm{d} = (d_1, \dots, d_{100}) \sim \mathcal{N}_{100}(\bm{0}, \sigma_{\mathrm{a}}^2 \bm{D})$, and
$\bm{x} = (x_1, \dots, x_{100}) \sim \mathcal{N}_{100}(\bm{0}, \sigma_{\mathrm{a}}^2 \bm{X})$, and we use a
sum-to-zero constraint on all genetic effects.
The covariance matrices $\bm{A}$, $\bm{D}$ and $\bm{X}$ are computed from the 
single nucleotide polymorphism (SNP) matrix with thousands of genetic markers, 
see \citet{hem2021} for details. 
This model has structured covariance matrices, and we use the
\code{"generic0"} latent model. 
\code{"generic0"} requires the argument
\code{Cmatrix}, which is the precision (inverse covariance) matrix
$\bm{Q}_*$ for
the effect.
With these data,
we get the following formula:
\begin{CodeChunk}
\begin{CodeInput}
R> formula <- y ~
+    mc(a, model = "generic0", Cmatrix = Q_a, constr = TRUE) +
+    mc(d, model = "generic0", Cmatrix = Q_d, constr = TRUE) +
+    mc(x, model = "generic0", Cmatrix = Q_x, constr = TRUE)
\end{CodeInput}
\end{CodeChunk}
We go through the reasoning behind a prior where we use all available 
prior knowledge, following the tree structure in Table \ref{tab:ex:genetic}.

\begin{table}[t!]
\centering
\begin{tabular}{cl}
	\textbf{Tree structure} & \textbf{Parameters, priors} \\ \hline \\[-8pt]
	\begin{tabular}{@{}c@{}} \tikz \pic{genetic_tree3}; \end{tabular} 	& 	
	\begin{tabular}{@{}l@{}}
	    $\sigma_{\mathrm{a+d+x}+\varepsilon}^2 \sim
		    \mathrm{Jeffreys'}$ \\[5pt]
		$\omega_{\frac{\mathrm{g}}{\mathrm{g} + \varepsilon}} \sim
		    \mathrm{PC}_{\mathrm{0}}(0.25)$ \\[5pt]
		$\omega_{\frac{\mathrm{a}}{\mathrm{g}}} \sim
		    \mathrm{PC}_{\mathrm{M}}(0.85, 0.8)$ \\[5pt]
		$\omega_{\frac{\mathrm{d}}{\mathrm{d+x}}} \sim
		    \mathrm{PC}_{\mathrm{M}}(0.67, 0.8)$ \\ \\[-8pt]
	\end{tabular} \\ \hline \\[-8pt]
\end{tabular}
\caption{
Tree structures and the corresponding parameters for the genomic example: $g_i = a_i + d_i + x_i$.
}
\label{tab:ex:genetic}
\end{table}

The expert in genetics has information on the heritability, which is the amount of total
variance attributed to the genetic effects and on the distribution of the genetic effect $\bm{g}$
to the additive, dominance and epistasis effects $\bm{a}$, $\bm{d}$ and $\bm{x}$.
The expert says the heritability
$\omega_{\frac{\mathrm{g}}{\mathrm{g}+\varepsilon}}$
is around 0.25, and that we want to avoid overfitting, which leads us to
$\omega_{\frac{\mathrm{g}}{\mathrm{g}+\varepsilon}} \sim \mathrm{PC}_{\mathrm{0}}(0.25)$.
The additive, dominance and epistasis effects have according to the expert a division of the genetic variance 
that
is around (85, 10, 5)\%, respectively. To achieve this, we must use two dual-splits to decompose the
genetic variation, and do this by splitting off the additive effect first, with a
$\mathrm{PC}_{\mathrm{M}}(0.85, 0.8)$ prior on $\omega_{\frac{\mathrm{a}}{\mathrm{g}}}$
(the amount of genetic variance that is additive).
Then we attribute the remaining 15\% of the genetic variance to $\bm{d}$ and $\bm{x}$ with
67\% to $\bm{d}$ with 
$\mathrm{PC}_{\mathrm{M}}(0.67, 0.8)$ on $\omega_{\frac{\mathrm{d}}{\mathrm{d+x}}}$.
We choose a concentration parameter value of $0.8$ because the expert is quite sure about
the (85, 10, 5)\% division. This corresponds to having 75\% of the density mass in
the interval $[\mathrm{logit}(m)-1, \mathrm{logit}(m)+1]$.
The expert does not want to use expert knowledge for the total variance
$\sigma_{\mathrm{a+d+x}+\varepsilon}^2$,
so we use Jeffreys' prior.

\begin{figure}
    \centering
    \includegraphics[width = 0.8\linewidth]{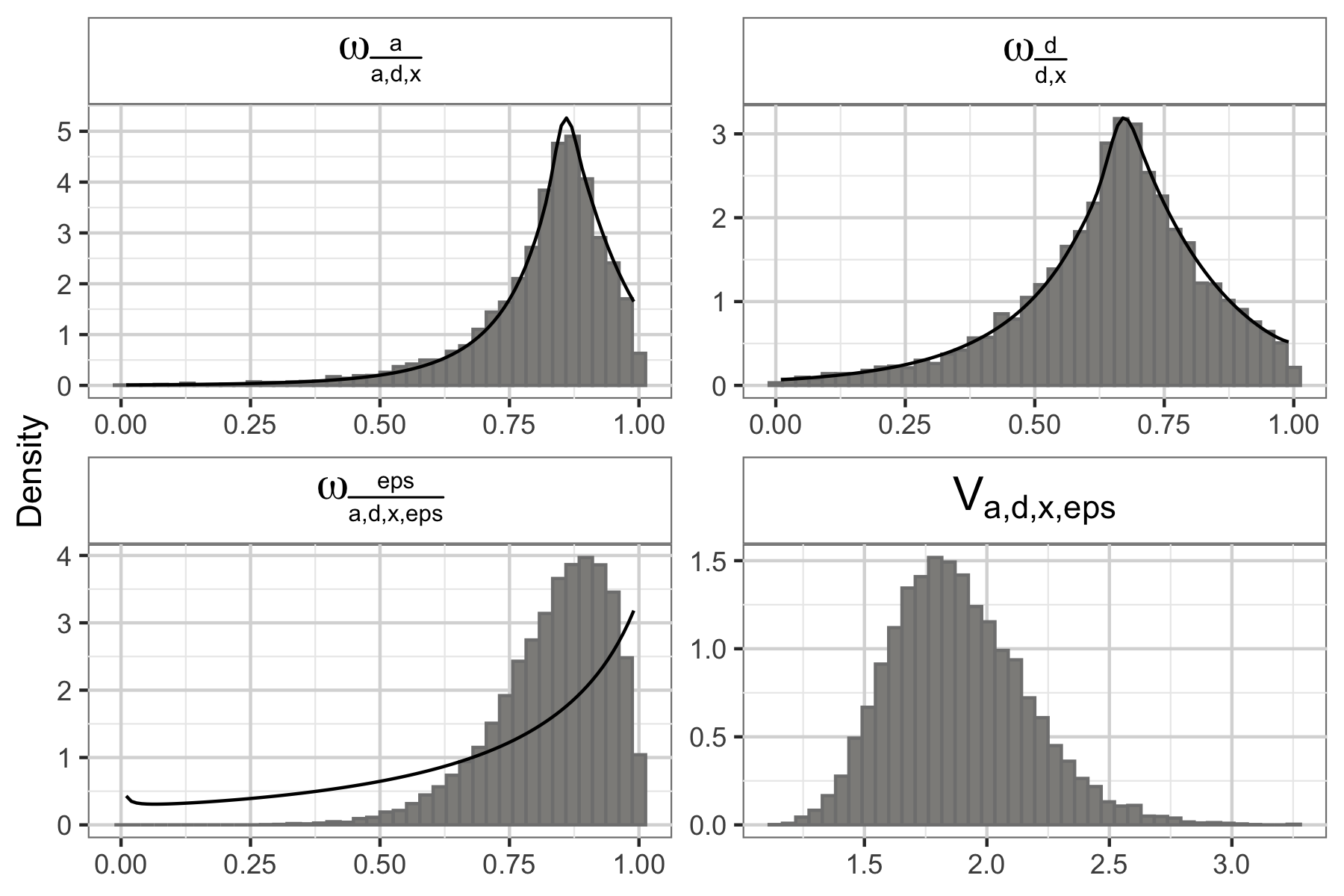}
    \caption{Prior and posterior distribution of the random effect parameters for the genomic selection example.}
    \label{fig:ex:genetic}
\end{figure}

We have simulated a dataset following the description in \citet{hem2021}
\citep[see also][]{gaynor2017, selle2019}, using the \proglang{R} package
\pkg{AlphaSimR} \citep{faux2016, gaynor2019}. The source code for
simulating the dataset is available in the Supplemental Materials in
\citet{hem2021} \citep{hem2021supp}.
The dataset is included as \code{wheat\_data} in \pkg{makemyprior}. 
To incorporate the expert knowledge in a unified way, 
we first scale the covariance matrices have typical variance
equal to 1 (see \citet{sorbye-rue-2014} for details), using the function \fct{scale\_precmat} in 
\pkg{makemyprior}:
\begin{CodeChunk}
\begin{CodeInput}
R> wheat_data_scaled <- wheat_data
R> wheat_data_scaled$Q_a <- scale_precmat(wheat_data$Q_a)
R> wheat_data_scaled$Q_d <- scale_precmat(wheat_data$Q_d)
R> wheat_data_scaled$Q_x <- scale_precmat(wheat_data$Q_x)
\end{CodeInput}
\end{CodeChunk}
This model is implemented
as follows:
\begin{CodeChunk}
\begin{CodeInput}
R> prior <- make_prior(formula, wheat_data_scaled, prior = list(
+    tree = "s1 = (d, x); s2 = (a, s1); s3 = (s2, eps)",
+    w = list(s1 = list(prior = "pcM", param = c(0.67, 0.8)),
+             s2 = list(prior = "pcM", param = c(0.85, 0.8)),
+             s3 = list(prior = "pc0", param = 0.25))))
\end{CodeInput}
\end{CodeChunk}
%
Note that we omit the specification of the total variance prior, as
we choose the default Jeffreys' prior. It can be specified with adding
\code{V = list(s3 = list(prior = "jeffreys"))} to the list provided to the
\code{prior} argument.
We now do inference on this model and plot the results:
\begin{CodeChunk}
\begin{CodeInput}
R> posterior <- inference_stan(prior, iter = 15000, warmup = 5000,
+                              seed = 1, init = "0", chains = 1)
R> plot_posterior_stan(posterior, param = "prior", prior = TRUE)
\end{CodeInput}
\begin{CodeOutput}
Tree structure: d_x = (d,x); a_d_x = (a,d_x); eps_a_d_x = (eps,a_d_x)

Weight priors:
	w[d/d_x] ~ PCM(0.67, 0.8)
	w[a/a_d_x] ~ PCM(0.85, 0.8)
	w[eps/eps_a_d_x] ~ PC1(0.75)
Total variance priors:
	V[eps_a_d_x] ~ Jeffreys'


SAMPLING FOR MODEL 'full_file' NOW (CHAIN 1).
Chain 1: 
Chain 1: Gradient evaluation took 0.000232 seconds
Chain 1: 1000 transitions using 10 leapfrog steps per transition 
would take 2.32 seconds.
Chain 1: Adjust your expectations accordingly!
Chain 1: 
Chain 1: 
Chain 1: Iteration:     1 / 15000 [  0%]  (Warmup)
Chain 1: Iteration:  1500 / 15000 [ 10%]  (Warmup)
Chain 1: Iteration:  3000 / 15000 [ 20%]  (Warmup)
Chain 1: Iteration:  4500 / 15000 [ 30%]  (Warmup)
Chain 1: Iteration:  5001 / 15000 [ 33%]  (Sampling)
Chain 1: Iteration:  6500 / 15000 [ 43%]  (Sampling)
Chain 1: Iteration:  8000 / 15000 [ 53%]  (Sampling)
Chain 1: Iteration:  9500 / 15000 [ 63%]  (Sampling)
Chain 1: Iteration: 11000 / 15000 [ 73%]  (Sampling)
Chain 1: Iteration: 12500 / 15000 [ 83%]  (Sampling)
Chain 1: Iteration: 14000 / 15000 [ 93%]  (Sampling)
Chain 1: Iteration: 15000 / 15000 [100%]  (Sampling)
Chain 1: 
Chain 1:  Elapsed Time: 17.7655 seconds (Warm-up)
Chain 1:                42.6968 seconds (Sampling)
Chain 1:                60.4623 seconds (Total)
Chain 1: 
\end{CodeOutput}
\end{CodeChunk}
Figure \ref{fig:ex:genetic} shows the prior and posterior together on the parameterization of
the prior (see Table \ref{tab:ex:genetic}).


We see that 
we do not have enough data to estimate the variance proportion
for the additive and nonadditive genetic effects: The posterior distribution is
almost identical to the prior distribution.
\citet{hem2021} have conducted an extensive simulation study on this and similar
models. They saw a strong need for robust prior distributions, which we
also see in Figure \ref{fig:ex:genetic}, because the nonadditive
effects $\bm{d}$ and $\bm{x}$ are strongly confounded with the environmental
effect $\bm{\varepsilon}$, and the number of observations is small compared to
the number of genetic markers that needs to be estimated \citep{sorensen2007}.

\subsubsection{Latin square experiment}

We consider the latin square experiment
in Example \ref{exmp:doe}.
In line with \citet[][Section 5.2]{fuglstad2020},
we expand the model and assume the treatment effect now consists of a smooth signal
$\bm{c}^{(1)} = (c_1^{(1)}, \dots, c_9^{(1)}) \sim
(\bm{0}, \sigma_{\mathrm{c}^{(1)}}^2 \mathbf{Q}_{\mathrm{RW2}}^{-1})$ where
$\sigma_{\mathrm{c}^{(1)}}^2$ is the variance and $\mathbf{Q}_{\mathrm{RW2}}^{-1}$ is the covariance
matrix describing the intrinsic second-order random walk
\citep[][Chapter 3]{gmrfbook},
and random noise
$\bm{c}^{(2)} = (c_1^{(2)}, \dots, c_9^{(2)})
\sim \mathcal{N}_9(\bm{0}, \sigma_{\mathrm{c}^{(2)}}^2 \mathbf{I}_9)$.
We remove implicit intercept and linear effect by requiring 
$\sum_{i=1}^9 c_i^{(1)} = 0$ and
$\sum_{i=1}^9 i c_i^{(1)} = 0$.
To simplify the notation, we use 
$f_{i,j} = a_i + b_j + c_{k[i,j]}^{(1)} + c_{k[i,j]}^{(2)}$.

We show how to create the prior distributions in Table \ref{tab:ex:latin}.
We want to avoid overfitting of the model, and use a prior with shrinkage towards
the residuals in the top split with a median giving 75\% residual effect. 
We do not have any preference for the attribution of the row, column and treatment effects,
and use an ignorant Dirichlet prior
for the middle split.
In the bottom split we
again we want to avoid overfitting, and use a prior with shrinkage towards the unstructured treatment
effect and a median corresponding to 75\% unstructured treatment effect.
At last we do not want to say anything about the scale of the total variance,
and use the default Jeffreys' prior.

\begin{table}[t!]
\centering
\begin{tabular}{cl}
	\textbf{Tree structure} & \textbf{Parameters, priors} \\ \hline \\[-8pt]
	\begin{tabular}{@{}c@{}} \tikz \pic{latin_tree3}; \end{tabular} 	& 	
	\begin{tabular}{@{}l@{}}
	    $\sigma_{\mathrm{a+b} + \mathrm{c}^{(1)} + \mathrm{c}^{(2)} + \varepsilon}^2 \sim 
	    \mathrm{Jeffreys'}$ \\[5pt]
	    $\omega_{\frac{f_i}{f_i + \varepsilon}} \sim 
	    \mathrm{PC}_{\mathrm{0}}(0.25)$ \\[5pt]
	    $\big(\omega_{\frac{\mathrm{a}}{f_i}},
	    \omega_{\frac{\mathrm{b}}{f_i}}, 1-\omega_{\frac{\mathrm{a}}{f_i}} -
	    \omega_{\frac{\mathrm{b}}{f_i}} \big) \sim
	    \mathrm{Dirichlet}(3)$ \\[5pt]
	    $\omega_{\frac{\mathrm{c}^{(1)}}{\mathrm{c}^{(1)} +\mathrm{c}^{(2)}}} \sim
	    \mathrm{PC}_{\mathrm{0}}(0.25)$ \\ \\[-8pt]
    \end{tabular} \\ \hline \\[-8pt]
\end{tabular}
\caption{
Tree structures and the corresponding parameters for the prior used in the latin square model:
$f_{i,j} = a_i + b_j + c_{k[i,j]}^{(1)} + c_{k[i,j]}^{(2)}$.
}
\label{tab:ex:latin}
\end{table}

\begin{figure}[t]
    \centering
    \includegraphics[width = 1\linewidth]{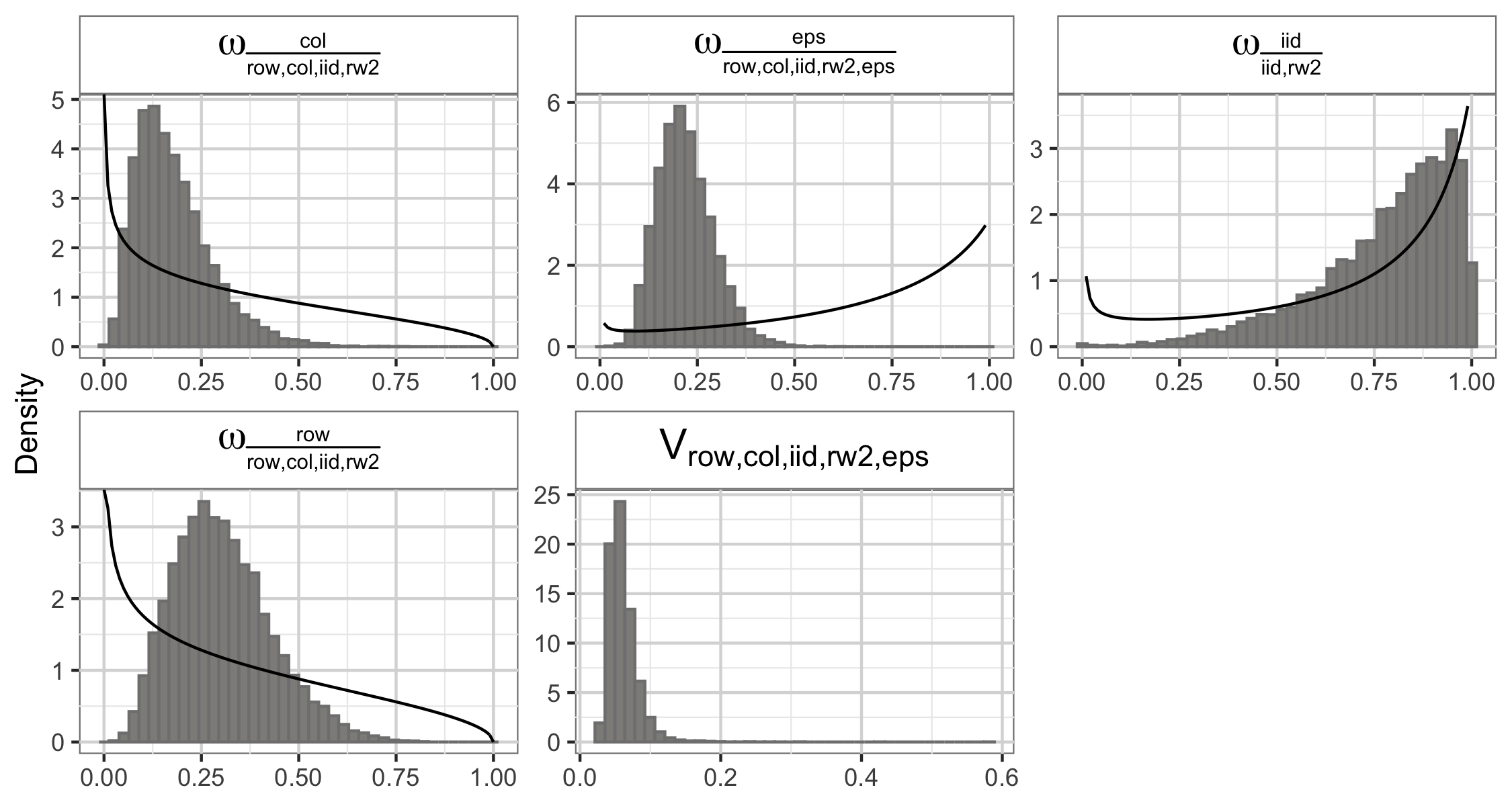}
    \caption{Prior and posterior distribution of the random effect parameters for the latin square example.}
    \label{fig:ex:latin}
\end{figure}

The dataset used in this model is included in \pkg{makemyprior} as \code{latin\_data}.
It is a simulated dataset, following the description of \citet[][Section 5.2]{fuglstad2020}, where we have used
$\sigma_{\mathrm{a}} = \sigma_{\mathrm{b}} = \sigma_{\mathrm{c}^{(2)}}
= \sigma_{\varepsilon} = 0.1$ and true treatment effect
$c_i^{(1)} = 0.02 \cdot ((i-5)^2 - 20/3)$.
\code{lin} in the formula below is the linear effect of treatment 
$k[i,j]$ and both the intercept $\mu$ and the coefficient $\beta$ has a default 
$\mathcal{N}(0, \sigma = 1000)$ prior.
The following will fit this model and produce the plots in Figure \ref{fig:ex:latin}:
\begin{CodeChunk}
\begin{CodeInput}
R> formula <- y ~ lin + mc(row) + mc(col) + mc(iid) +
+    mc(rw2, model = "rw2", constr = TRUE, lin_constr = TRUE)
R> prior <- make_prior(
+    formula, latin_data,
+    prior = list(tree = "s1 = (rw2, iid);
+                                 s2 = (row, col, s1); s3 = (s2, eps)",
+                 w = list(s1 = list(prior = "pc0", param = 0.25),
+                          s2 = list(prior = "dirichlet"),
+                          s3 = list(prior = "pc0", param = 0.25))))

R> posterior <- inference_stan(prior, iter = 15000, warmup = 5000,
+                              seed = 1, init = "0", chains = 1,
+                              control = list(adapt_delta = 0.9))
R> plot_posterior_stan(posterior, param = "prior", prior = TRUE)
\end{CodeInput}
\begin{CodeOutput}
Tree structure: iid_rw2 = (iid,rw2); row_col_iid_rw2 = (row,col,iid_rw2); 
eps_row_col_iid_rw2 = (eps,row_col_iid_rw2)

Weight priors:
	w[iid/iid_rw2] ~ PC1(0.75)
	(w[row/row_col_iid_rw2], w[col/row_col_iid_rw2]) ~ Dirichlet(3)
	w[eps/eps_row_col_iid_rw2] ~ PC1(0.75)
Total variance priors:
	V[eps_row_col_iid_rw2] ~ Jeffreys'


SAMPLING FOR MODEL 'full_file' NOW (CHAIN 1).
Chain 1: 
Chain 1: Gradient evaluation took 0.000262 seconds
Chain 1: 1000 transitions using 10 leapfrog steps per transition 
would take 2.62 seconds.
Chain 1: Adjust your expectations accordingly!
Chain 1: 
Chain 1: 
Chain 1: Iteration:     1 / 15000 [  0%]  (Warmup)
Chain 1: Iteration:  1500 / 15000 [ 10%]  (Warmup)
Chain 1: Iteration:  3000 / 15000 [ 20%]  (Warmup)
Chain 1: Iteration:  4500 / 15000 [ 30%]  (Warmup)
Chain 1: Iteration:  5001 / 15000 [ 33%]  (Sampling)
Chain 1: Iteration:  6500 / 15000 [ 43%]  (Sampling)
Chain 1: Iteration:  8000 / 15000 [ 53%]  (Sampling)
Chain 1: Iteration:  9500 / 15000 [ 63%]  (Sampling)
Chain 1: Iteration: 11000 / 15000 [ 73%]  (Sampling)
Chain 1: Iteration: 12500 / 15000 [ 83%]  (Sampling)
Chain 1: Iteration: 14000 / 15000 [ 93%]  (Sampling)
Chain 1: Iteration: 15000 / 15000 [100%]  (Sampling)
Chain 1: 
Chain 1:  Elapsed Time: 20.9547 seconds (Warm-up)
Chain 1:                44.9349 seconds (Sampling)
Chain 1:                65.8897 seconds (Total)
Chain 1: 
\end{CodeOutput}
\end{CodeChunk}
Figure \ref{fig:ex:latin} shows the prior and posterior together on the parameterization of
the prior.
The posterior distribution of the bottom split, 
attributing the treatment effect to the
random noise and smooth signal, is only slightly different from the prior, 
indicating that there is no strong signal about the smooth treatment
effect in the data. By using a prior with shrinkage towards only
random noise treatment effect, we avoid overfitting.
The model has learned about the three other variance
proportions, and we see that even though the prior on the amount of
total variance going to the residual effect has shrinkage towards $1$, the
model is not restricted by this (top right plot).

Note that for inference with INLA, we must implement the linear constraint with 
\code{extraconstr = list(A = matrix(1:9, 1, 9), e = matrix(0, 1, 1))} 
in \code{mc(rw2, ...)} in the formula.


\subsection{Binomial responses: Neonatal mortality}
\label{sec:ex:neonatal}

This example is based on a study carried out by \citet{fuglstad2020}.
Child mortality is an important indicator of health and well-being in a country.
We define neonatal mortality as the number of deaths of infants the first month of life per live birth,
which can be estimated using national household surveys from 
Demographic and Health Surveys \citep{dhskenya}.
From such surveys we can extract 
the number of live births $b_{i,j}$ and
the number of neonatal deaths $y_{i,j}$ 
in cluster $j$ in county $i$,
and use an indicator $x_{i,j}$ for classifying cluster $j$ in county $i$ as rural ($x_{i,j} = 0)$ 
or urban ($1$).
We model $y_{i,j} | b_{i,j}, p_{i,j} \sim \mathrm{Binomial}(b_{i,j}, p_{i,j})$ with
the linear predictor
\begin{equation}
    \eta_{i,j} = \mathrm{logit}(p_{i,j}) = 
    \mu + x_{i,j} \beta + u_i + v_i + \nu_{i,j}, \
    i = 1, \dots, n, \ j = 1, \dots, m_i,
\end{equation}
where $v_i \sim \mathcal{N}(0, \sigma_v^2)$ and 
$\nu_{i,j} \sim \mathcal{N}(0, \sigma_{\nu}^2)$ are i.i.d. random effects
with sum-to-zero constraints
for counties and clusters, respectively,
and $\bm{u}$ is a Besag effect on county
with variance $\sigma_u^2$ and a sum-to-zero constraint.
In the Besag model, the spatial effect of each county depends on the effects
in the neighboring regions (see e.g. \citet{besag1991} for details),
and when combining it with an i.i.d. effect on the same level in the hierarchy,
we get a BYM (Besag, York and Mollié) model \citep{besag1991}.
We want to investigate whether or not there is a spatial effect present.

We simulated a dataset following the description in
\citet[][Section 6.2]{fuglstad2020} with the 47 counties in Kenya
(see Figure \ref{fig:ex:neonatal_map} for a map).
We used 6, 7 or 8 clusters in each county
which gave in total $327$ clusters, and thus $327$ observations,
$b_{i,j} = 25$ live births in each cluster, 
and parameters
$\mu = -4$,
$\beta = 0.1$,
$\sigma_{\nu}^2 = 0.2$,
$\sigma_{\mathrm{v}}^2 = 0.1$, and
$\sigma_{\mathrm{u}}^2 = 0.5$.

This dataset is available in \pkg{makemyprior} as \code{neonatal\_data}, 
together with
other necessary files for fitting the model.

\begin{table}[t!]
\centering
\begin{tabular}{cl}
	\textbf{Tree structure} & \textbf{Parameters, priors} \\ \hline 
	\\[-8pt]
	\begin{tabular}{@{}c@{}} \tikz \pic{kenya_tree1}; \end{tabular} & 	
	\begin{tabular}{@{}l@{}}
							$\sigma_{\mathrm{u + v} + \nu}^2 \sim
							\mathrm{PC}_0(3.35, 0.05)$ \\[5pt]
							$\omega_{\frac{\mathrm{u+v}}{\mathrm{u+v}+\nu}} \sim
							\mathrm{PC}_{\mathrm{1}}(0.75)$ \\[5pt]
							$\omega_{\frac{\mathrm{u}}{\mathrm{u+v}}} \sim
							\mathrm{PC}_{\mathrm{0}}(0.25)$ \\ \\[-8pt]
							\end{tabular} \\ \hline 
							\\[-8pt]
\end{tabular}
\caption{
Tree structures and the corresponding parameters for the neonatal mortality model.
}
\label{tab:ex:neonatal}
\end{table}

\begin{figure}[t]
    \centering
    \begin{subfigure}[t]{0.63\textwidth}
        \centering
        \includegraphics[width = 1\linewidth]{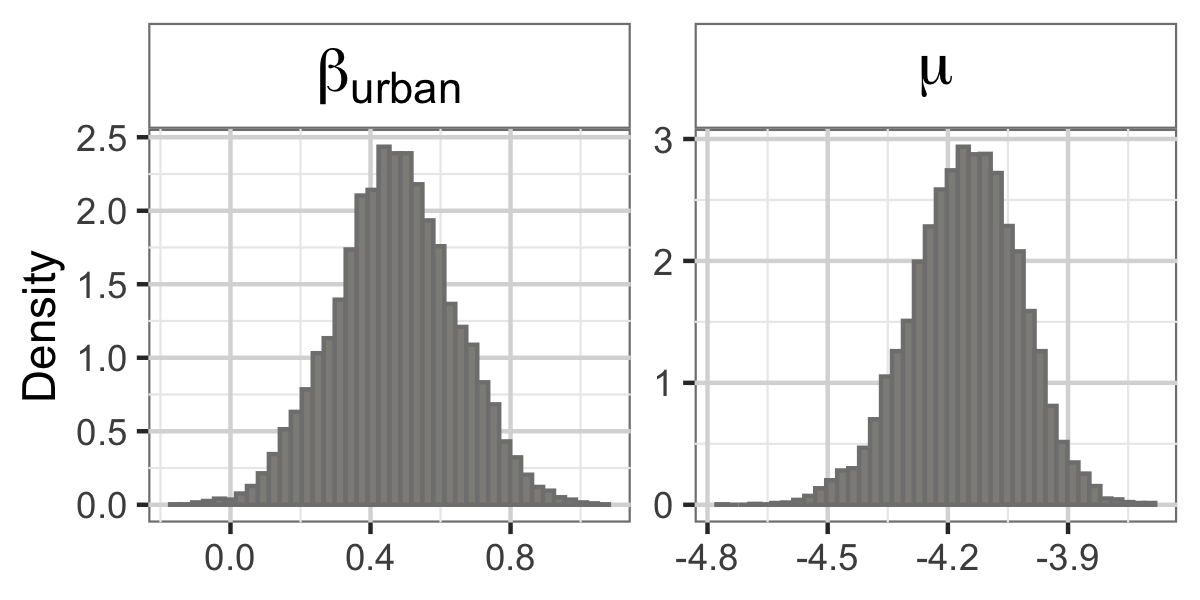}
        \caption{Posterior of the effect of urban/rural and the intercept.}
        \label{fig:ex:neonatal:fix}
    \end{subfigure}
    
    \begin{subfigure}[t]{1\textwidth}
        \centering
        \includegraphics[width = 1\linewidth]{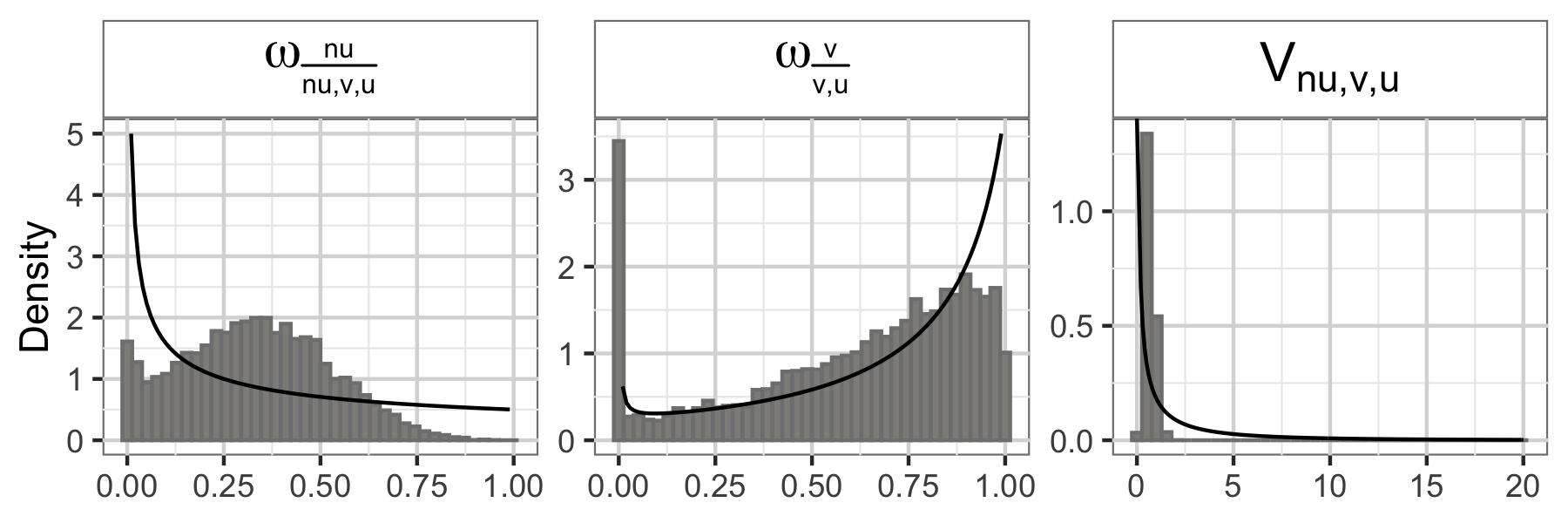}
        \caption{Random effect parameters.}
        \label{fig:ex:neonatal:ran}
    \end{subfigure}
    \caption{Prior and posterior distribution of
    \subref{fig:ex:neonatal:fix}) coefficients of the fixed effects and
    \subref{fig:ex:neonatal:ran}) total variance and variance proportions
    of the random effects 
    for the neonatal mortality example.}
    \label{fig:ex:neonatal}
\end{figure}

\begin{figure}[p]
    \centering
    \includegraphics[width = 0.5\linewidth]{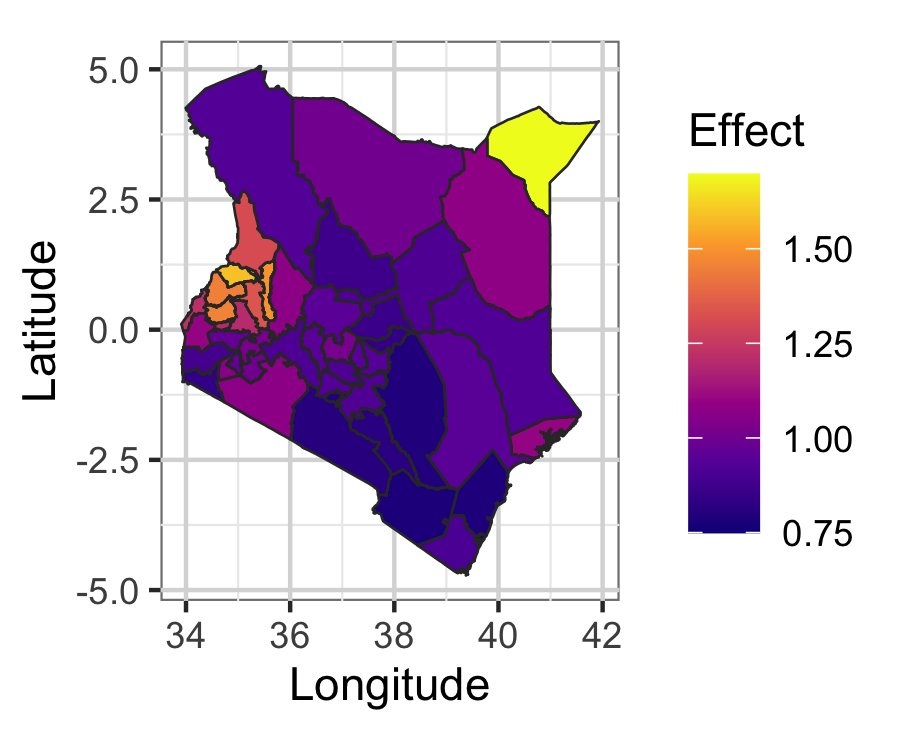}
    \caption{Posterior median of $e^{u_i}$ for each county in Kenya. Note
    that this is based on simulated data.}
    \label{fig:ex:neonatal_map}

    \centering
    \includegraphics[width = 1\linewidth]{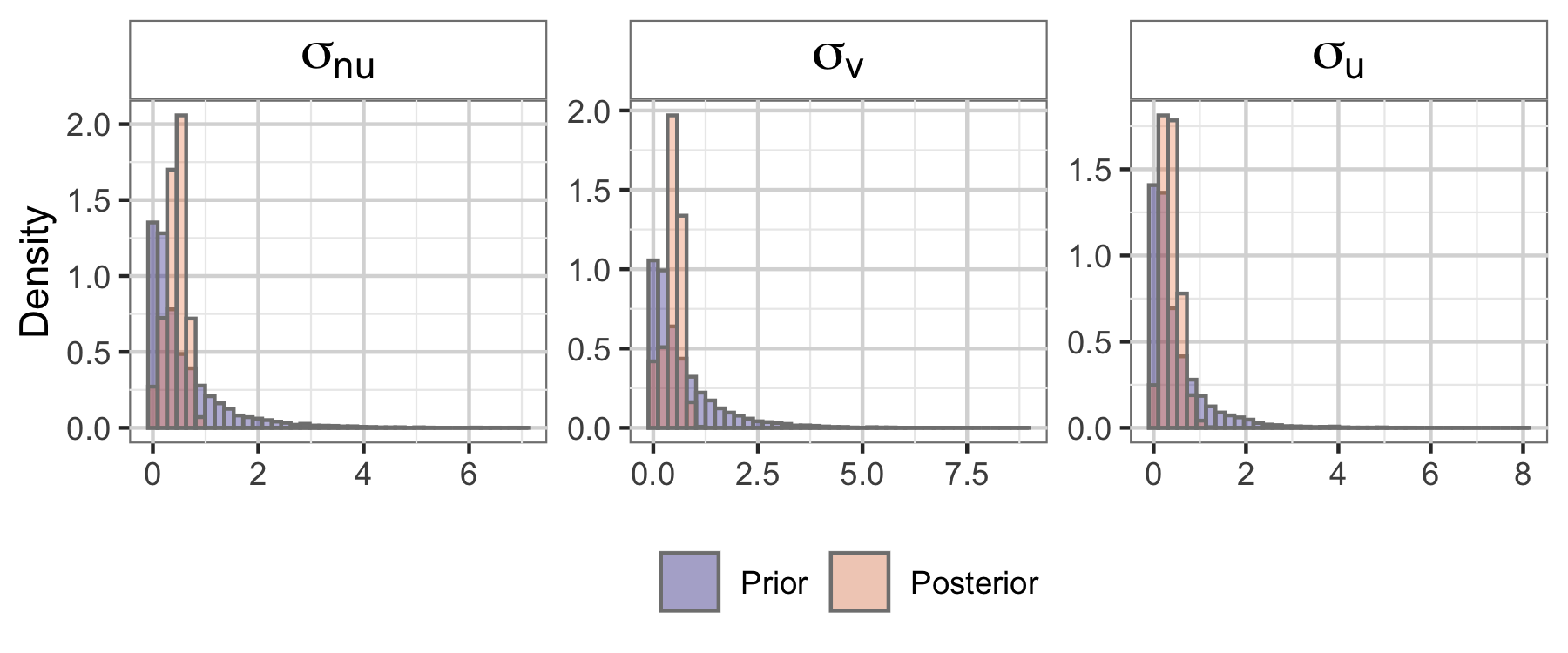}
    \caption{Prior and posterior distribution for the neonatal mortality example
    on standard deviation scale.}
    \label{fig:ex:neonatal_stdev}
\end{figure}

We prefer coarser over finer unstructured effects, and unstructured over structured effects.
That means that we prefer $\bm{v}$ over $\bm{u}$ and $\bm{v} + \bm{u}$ over $\bm{\nu}$ in
the prior.
The BYM model is intuitively represented with a dual split
in the prior tree, where one leaf node represents a
Besag effect and the other represents an i.i.d. effect.
We achieve this with a prior that distributes the between-county variance with shrinkage towards 
the unstructured county effect,
which gives the BYM2 model of \citet{riebler2016},
and we shrink the total variance towards the county effects.
Following \citet{fuglstad2020}, we induce shrinkage on the total variance such that we have a 
90\% credible interval of
$(0.1, 10)$ for the effect of $\exp(v_i + u_i + \nu_{i,j})$. We use 
the function \fct{find\_pc\_prior\_param}
in \pkg{makemyprior} to find the parameters for the PC prior:
\begin{CodeChunk}
\begin{CodeInput}
R> set.seed(1)
R> find_pc_prior_param(lower = 0.1, upper = 10, prob = 0.9, N = 2e5)
\end{CodeInput}
\begin{CodeOutput}
U = 3.353132
Prob(0.09866969 < exp(eta) < 9.892902) = 0.9
\end{CodeOutput}
\end{CodeChunk}
This gives a $\mathrm{PC}_{\mathrm{0}}(3.35, 0.05)$ prior.
The tree structure and a summary of the prior distributions can be found in Table
\ref{tab:ex:neonatal}. We fit the model with \proglang{Stan}:
\begin{CodeChunk}
\begin{CodeInput}
R> graph_path <- paste0(path.package("makemyprior"), "/neonatal.graph")
R> formula <- y ~ mc(nu) + mc(v) + 
+    mc(u, model = "besag", graph = graph_path, scale.model = TRUE)
R> prior <- make_prior(
+      formula, neonatal_data, family = "binomial",
+      prior = list(tree = "s1 = (u, v); s2 = (s1, nu)",
+                   w = list(s1 = list(prior = "pc0", param = 0.25),
+                            s2 = list(prior = "pc1", param = 0.75)),
+                   V = list(s2 = list(prior = "pc", 
+                                      param = c(3.35, 0.05)))))
R> posterior <- inference_stan(prior, iter = 15000, warmup = 5000, 
+                              seed = 1, init = "0", chains = 1,
+                              control = list(adapt_delta = 0.9))
\end{CodeInput}
\begin{CodeOutput}
Tree structure: v_u = (v,u); nu_v_u = (nu,v_u)

Weight priors:
	w[v/v_u] ~ PC1(0.75)
	w[nu/nu_v_u] ~ PC0(0.25)
Total variance priors:
	sqrt(V)[nu_v_u] ~ PC0(3.35, 0.05)


SAMPLING FOR MODEL 'full_file' NOW (CHAIN 1).
Chain 1: 
Chain 1: Gradient evaluation took 0.000272 seconds
Chain 1: 1000 transitions using 10 leapfrog steps per transition 
would take 2.72 seconds.
Chain 1: Adjust your expectations accordingly!
Chain 1: 
Chain 1: 
Chain 1: Iteration:     1 / 15000 [  0%]  (Warmup)
Chain 1: Iteration:  1500 / 15000 [ 10%]  (Warmup)
Chain 1: Iteration:  3000 / 15000 [ 20%]  (Warmup)
Chain 1: Iteration:  4500 / 15000 [ 30%]  (Warmup)
Chain 1: Iteration:  5001 / 15000 [ 33%]  (Sampling)
Chain 1: Iteration:  6500 / 15000 [ 43%]  (Sampling)
Chain 1: Iteration:  8000 / 15000 [ 53%]  (Sampling)
Chain 1: Iteration:  9500 / 15000 [ 63%]  (Sampling)
Chain 1: Iteration: 11000 / 15000 [ 73%]  (Sampling)
Chain 1: Iteration: 12500 / 15000 [ 83%]  (Sampling)
Chain 1: Iteration: 14000 / 15000 [ 93%]  (Sampling)
Chain 1: Iteration: 15000 / 15000 [100%]  (Sampling)
Chain 1: 
Chain 1:  Elapsed Time: 39.9314 seconds (Warm-up)
Chain 1:                134.344 seconds (Sampling)
Chain 1:                174.276 seconds (Total)
Chain 1: 
\end{CodeOutput}
\end{CodeChunk}
For inference with INLA, the \code{Ntrials} argument must be provided
to \fct{inference\_inla}.
The following produce the plots in
Figure \ref{fig:ex:neonatal} and gives some key information on the posterior:
\begin{CodeChunk}
\begin{CodeInput}
R> plot_posterior_fixed(posterior)
R> plot_posterior_stan(posterior, param = "prior", prior = TRUE)
R> posterior
\end{CodeInput}
\begin{CodeOutput}
Model: y ~ urban + mc(nu) + mc(v) + mc(u, model = "besag", 
graph = graph_path, 
    scale.model = TRUE)
Tree structure: v_u = (v,u); nu_v_u = (nu,v_u) 

Inference done with Stan.

 Param.       mean   median sd   
 V[nu_v_u]     0.668  0.642 0.243
 w[v/v_u]      0.610  0.688 0.302
 w[nu/nu_v_u]  0.329  0.328 0.189
 intercept    -4.155 -4.151 0.135
 urban         0.470  0.469 0.171
\end{CodeOutput}
\end{CodeChunk}
Figure \ref{fig:ex:neonatal_map} shows the posterior spatial effect $e^{u_i}$ plotted in a map.
We see a spatial variation between the counties.
The necessary
data for creating the spatial map are not included, but can be obtained from
\url{https://gadm.org/}.
The samples for the effects can easily be extracted
with \fct{extract\_posterior\_effect}
which take the arguments \code{obj} from
\fct{inference\_stan} and the name of the effect:
\begin{CodeChunk}
\begin{CodeInput}
R> u <- extract_posterior_effect(posterior, "u")
\end{CodeInput}
\end{CodeChunk}
%

The fixed effects in Figure \ref{fig:ex:neonatal:fix} show that the
intercept is not contributing much to the linear predictor, while 
the effect of urban/rural shows that there is a higher mortality in urban areas
(which is the case also for real data, see \citet{dhskenya}).
From Figure \ref{fig:ex:neonatal:ran} we see that the model
has learned about the total variance from the data and
about
the amount of total (latent) variance to the cluster effect ($\bm{\nu}$), but 
there is not enough information in the data
about the
amount of county variance to the structured county effect ($\bm{u}$). 
The
following 
fits the model without the likelihood (sampling from the prior) and
produces the plots of the prior and posterior on standard deviation scale
in Figure \ref{fig:ex:neonatal_stdev}:
\begin{CodeChunk}
\begin{CodeInput}
R> prior_samps <- inference_stan(prior, use_likelihood = F, print_prior = F,
+                                iter = 15000, warmup = 5000,
+                                seed = 1, init = "0", chains = 1)
R> plot_several_posterior_stan(list(Prior = prior_samps, 
+                                   Posterior = posterior), "stdev")
\end{CodeInput}
\begin{CodeOutput}
SAMPLING FOR MODEL 'full_file' NOW (CHAIN 1).
Chain 1: 
Chain 1: Gradient evaluation took 0.000105 seconds
Chain 1: 1000 transitions using 10 leapfrog steps per transition 
would take 1.05 seconds.
Chain 1: Adjust your expectations accordingly!
Chain 1: 
Chain 1: 
Chain 1: Iteration:     1 / 15000 [  0%]  (Warmup)
Chain 1: Iteration:  1500 / 15000 [ 10%]  (Warmup)
Chain 1: Iteration:  3000 / 15000 [ 20%]  (Warmup)
Chain 1: Iteration:  4500 / 15000 [ 30%]  (Warmup)
Chain 1: Iteration:  5001 / 15000 [ 33%]  (Sampling)
Chain 1: Iteration:  6500 / 15000 [ 43%]  (Sampling)
Chain 1: Iteration:  8000 / 15000 [ 53%]  (Sampling)
Chain 1: Iteration:  9500 / 15000 [ 63%]  (Sampling)
Chain 1: Iteration: 11000 / 15000 [ 73%]  (Sampling)
Chain 1: Iteration: 12500 / 15000 [ 83%]  (Sampling)
Chain 1: Iteration: 14000 / 15000 [ 93%]  (Sampling)
Chain 1: Iteration: 15000 / 15000 [100%]  (Sampling)
Chain 1: 
Chain 1:  Elapsed Time: 3.38377 seconds (Warm-up)
Chain 1:                11.5259 seconds (Sampling)
Chain 1:                14.9097 seconds (Total)
Chain 1: 
\end{CodeOutput}
\end{CodeChunk}
From these graphs we see that the posterior of the standard deviations are
clearly different from the prior. 
We saw in Figure \ref{fig:ex:neonatal:ran} 
that the model did not learn much about the amount of county variation 
accounted for by the Besag effect ($\bm{u}$), 
but we cannot see this
from plots of the posterior standard deviations, and they do not show the whole picture.
This is another advantage of the HD prior: it is easy to see that 
even though we get the impression that the model has learned from the data,
that knowledge is not necessarily about the whole model.
This shows, as \citet{fuglstad2020} points out, that one should be
careful before drawing conclusions on first impressions about the results,
and more investigation should be done.

\section{Summary and discussion}
\label{sec:summary}

The \pkg{makemyprior} package offers an intuitive and transparent way of choosing and visualizing
prior distributions for Bayesian hiearchical models. It is easy to utilize expert knowledge, and clear what prior distributions are
used, also when the default settings are chosen. 
The package works with the flexible and widely used latent Gaussian models, and
offers Gaussian, binomial and Poisson likelihoods. This, together with the 
latent models included, 
makes the package applicable in a range of problems and applications.


The package can be used to investigate the prior choices, and
the user can simulate from the prior with \proglang{Stan} 
and look at the prior distributions
on different parameterizations. In this way, crucial misunderstandings of what prior distributions
are used can be discovered and corrected, and thus increase the understanding and
meaning of the prior.
The usage of the hierarchical decomposition (HD) prior 
is not limited to inference carried out with
\pkg{rstan} or \pkg{INLA}. Joint priors can be constructed and evaluated using \pkg{makemyprior} for use in external inference code or potentially incorporated in other \proglang{R} packages.

To see how the individual fixed effects contribute to the total data variation
would be interesting, and could be done with the HD prior framework.
However,
fixed effects are often correlated, and the variance that is explained
by each single fixed effect is not well defined.
The perhaps most intuitive way to include fixed effects
directly in the tree structure is to assign one
variance parameter to each effect, but
this can quickly increase the amount of variance parameters
to a level where inference become computationally hard.
\citet{gelmanbook} and \citet{zhang2020} have proposed prior distributions related to
the coefficient of determination, $R^2$, which measures
the amount of variance explained by the model. The generalized
$R^2$ proposed by \citet{gelmanbook} measures this at each level in the hierarchical
model. \citet{r2d2} extended the framework by \citet{zhang2020} to generalized linear mixed models. 
To include fixed effects directly in the joint prior in the HD prior framework 
is discussed by \citet{fuglstad2020}. We have chosen to
give them independent and vague priors.

Other exiting additions to the HD prior framework include
extending it to models outside the class of latent Gaussian models,
or to models where the hyperparameters of the priors will get prior distributions. This
will 
require further development of the framework, and may be highly computational expensive,
as the penalized complexity (PC) prior cannot be pre-computed in the same way as we do now with the
conditioning on the hyperparameters.

To open for using Dirichlet distributions with custom hyperparameter values
can be a natural next step. This addition to the HD prior framework itself
will open for even easier integration of the HD prior in other software (such as
Template Model Builder 
\citep[\pkg{TMB},][]{tmb}) as expert knowledge can be included without having to 
compute the PC prior. 
However, this will complicate the intuition behind the prior,
and it will be more difficult to use prior and expert knowledge in a 
transparent way. It
will require more thoughtful prior choices, and we lose one of the 
big advantages with the
easy-to-use and intuitive way of making priors with \pkg{makemyprior},
in addition to the shrinkage properties of the PC prior.
Allowing the user to specify custom prior distributions for variance parameters
will introduce even more flexibility, however, the included variance
prior distributions cover the most popular choices.

Including more latent models will 
further increase the amount of applications the package can be used for without 
specifying
custom \proglang{Stan}-code. This includes handling parameters such as correlations, which
can be done by giving them independent priors and conditioning on a 
representative value (e.g. the mean or median of the chosen prior) 
for this parameter when the joint prior is computed. For prior elicitation of these parameters additional \proglang{R}-packages might be useful. For example, the \proglang{R}-package \pkg{meta4diag} \citep{meta4diag}, which considers the analysis of diagnostic test studies, offers three strategies to intuitively
define a penalized complexity prior for a correlation parameter $\rho$ in a bivariate model given an arbitrary base value $\rho_0$, while the \pkg{INLA} package implements PC priors for autoregressive models \citep{sorbye2017}, see functions \code{pc.cor0} and \code{pc.cor1}. 
To open for easy integration into other software for inference, such as \pkg{TMB}, can be useful for models that are very complex
and will be highly time consuming and difficult 
to fit with \pkg{rstan} or \pkg{INLA}.

In conclusion, \pkg{makemyprior} offers something not offered by
the range of packages that can be used
to carry out inference for Bayesian hierarchical models. 
It makes it easy to include prior knowledge in an intuitive and transparent way,
can be used to verify prior choices, and
allows direct inference in a simple way.
\pkg{makemyprior} makes users aware of what priors are used and makes prior 
selection a concious choice, which is important
when doing inference 
to ensure that the model fitted is indeed the intended one.

\section*{Computational details}

The results in this paper were obtained with
\proglang{R} version 4.1.0 on platform
aarch64-apple-darwin20 (64-bit)
running under macOS Big Sur 11.3.1.

%
%
Package versions:
\pkg{makemyprior} 1.1.0,
\pkg{ggplot2} 3.3.5,
\pkg{Matrix} 1.3.3,
\pkg{methods} 4.1.0,
\pkg{shiny} 1.7.1,
\pkg{shinyjs} 2.1.0,
\pkg{shinyBS} 0.61,
\pkg{visNetwork} 2.1.0,
\pkg{rlang} 1.0.2,
\pkg{MASS} 7.3.54,
\pkg{rstan} 2.21.3,
\pkg{INLA} 21.11.22,
\pkg{knitr} 1.37,
and
\pkg{rmarkdown} 2.13.

%


\section*{Acknowledgments}

Hem, Fuglstad and Riebler were supported by project number 240873 from the Research Council of Norway.

\bibliography{references}

\newpage 

\begin{appendix}

\section[Code for Figure 2 in Section 3]
{Code for Figure \protect\NoHyper\ref{fig:pc_prior} in Section \protect\NoHyper\ref{sec:theory}}
\label{app:codefig2}

\begin{Code}
formula <- y ~ -1 + mc(a)
data <- list(
  a = rep(1:10, 10),
  y = rep(0, 100)
)

prior1 <- make_prior(
  formula, data, prior = list(tree = "(a); (eps)"))
prior2 <- make_prior(
  formula, data, 
  prior = list(tree = "s1 = (a, eps)", 
               w = list(s1 = list(prior = "pc0", param = 0.25))))
prior3 <- make_prior(
  formula, data, 
  prior = list(tree = "s1 = (a, eps)", 
               w = list(s1 = list(prior = "pc1", param = 0.75))))
prior4 <- make_prior(
  formula, data, 
  prior = list(tree = "s1 = (a, eps)", 
               w = list(s1 = list(prior = "pcM", param = c(0.25, 0.85)))))

plot_marginal_prior(seq(0, 5, 0.050), prior1, "sigma^2[a]", sd = T)
plot_marginal_prior(seq(0, 1, 0.001), prior2, "w[a/a_eps]")
plot_marginal_prior(seq(0, 1, 0.001), prior3, "w[a/a_eps]")
plot_marginal_prior(seq(0, 1, 0.001), prior4, "w[a/a_eps]")
\end{Code}

\end{appendix}

\end{document}